\documentclass[11pt]{article}

\usepackage[final]{acl}

\usepackage{times}
\usepackage{latexsym}

\usepackage[T1]{fontenc}

\usepackage[utf8]{inputenc}

\usepackage{microtype}

\usepackage{inconsolata}

\usepackage{graphicx}
\usepackage{tabularx}
\usepackage[table]{xcolor}
\usepackage{booktabs}
\usepackage{makecell}
\usepackage{array}
\usepackage{listings}
\usepackage{xcolor}
\usepackage{amssymb}
\usepackage{subcaption}
\usepackage{enumitem}
\usepackage{needspace}

\usepackage{placeins}

\usepackage{minted}
\setminted{
  fontsize=\footnotesize,
  breaklines,
  breakanywhere,
  autogobble,
  frame=single
}

\definecolor{codebg}{RGB}{255,252,242}
\definecolor{codekw}{RGB}{0,0,180}
\definecolor{codestr}{RGB}{163,21,21}
\definecolor{codecom}{RGB}{0,128,0}
\definecolor{codeflag}{RGB}{128,0,128}

\lstdefinestyle{mystyle}{
  basicstyle=\footnotesize\ttfamily,
  breaklines=true,
  frame=single,
  backgroundcolor=\color{codebg},
  showstringspaces=false,
  columns=fullflexible,
  keepspaces=true,
  keywordstyle=\color{codekw}\bfseries,
  stringstyle=\color{codestr},
  commentstyle=\color{codecom}\itshape,
  rulecolor=\color{black},
}

\lstdefinelanguage{json}{
  basicstyle=\footnotesize\ttfamily,
  showstringspaces=false,
  breaklines=true,
  keepspaces=true,
  string=[s]{"}{"},
  stringstyle=\color{codestr},
  comment=[l]{//},
  morecomment=[s]{/*}{*/},
  commentstyle=\color{codecom}\itshape,
  morekeywords={true,false,null},
  keywordstyle=\color{codekw}\bfseries
}

\lstdefinelanguage{bash}{
  alsoletter={-},
  morekeywords={
    sudo,apt,apt-get,cd,ls,cat,echo,mkdir,rm,cp,mv,
    python,python3,pip,pip3,export,chmod,source, --verbosity, --settings, --parallel
  },
  sensitive=true,
  morecomment=[l]{\#},
  morestring=[b]",
  morestring=[b]',
  literate=
    *{./tests/runtests.py}{{{\bfseries ./tests/runtests.py}}}1
}

\lstdefinelanguage{MyPython}{
  language=Python,
  morestring=[b]",
  morestring=[b]',
  morestring=[s]{"""}{"""},
  morestring=[s]{'''}{'''},
}

\usepackage{svg}

\usepackage{enumitem}

\title{ORACLE-SWE: Quantifying the Contribution of Critical Information Signals on SWE Agents\textsuperscript{\P}}

\author{
\textbf{Kenan Li\textsuperscript{1,*,\ensuremath{\dagger}}},
\textbf{Qirui Jin\textsuperscript{2,*,\ensuremath{\dagger}}},
 \textbf{Liao Zhu\textsuperscript{1}},
 \textbf{Xiaosong Huang\textsuperscript{1}},
 \textbf{Yijia Wu\textsuperscript{1}},
 \textbf{Yikai Zhang\textsuperscript{1}},
\\
 \textbf{Xin Zhang\textsuperscript{1}},
 \textbf{Zijian Jin \textsuperscript{1}},
 \textbf{Yufan Huang\textsuperscript{1}},
 \textbf{Elsie Nallipogu\textsuperscript{1}},
 \textbf{Chaoyun Zhang\textsuperscript{1}},
\\
 \textbf{Yu Kang\textsuperscript{1}},
 \textbf{Saravan Rajmohan\textsuperscript{1}},
 \textbf{Qingwei Lin\textsuperscript{1}},
 \textbf{Wenke Lee\textsuperscript{2}},
 \textbf{Dongmei Zhang\textsuperscript{1}}
\\
 \textsuperscript{1}Microsoft,
 \textsuperscript{2}Georgia Institute of Technology
}

\begin{document}
\maketitle

\begingroup
\renewcommand{\thefootnote}{\fnsymbol{footnote}}

\footnotetext[5]{
Accepted by \textbf{\textit{EACL 2027 Findings}}.
}

\footnotetext[1]{Equal contribution.}

\footnotetext[2]{
Correspondence to:
Kenan Li (\href{mailto:kenan.li.24@ucl.ac.uk}{kenan.li.24@ucl.ac.uk})
and Qirui Jin (\href{mailto:qiruijin@gatech.edu}{qiruijin@gatech.edu}).
}

\footnotetext[3]{
GitHub repo to extract \& evaluate the oracle \& agent-found signals from arbitrary SWE benchmarks following SWE-bench protocol: \href{https://github.com/njukenanli/ORACLE-SWE}{njukenanli/ORACLE-SWE}.
}

\footnotetext[4]{
The dataset of the extracted oracle signals from the three benchmarks: \href{https://huggingface.co/datasets/SWE-bench-Live/ORACLE-SWE}{SWE-bench-Live/ORACLE-SWE}.
}.

\endgroup

\begin{abstract}
Recent advances in Language Model (LM) agents have increasingly focused on repository-level software engineering (SWE) tasks, leading to various agentic workflows, training methods and failure analysis. Interestingly, rather than directly generating or evaluating edit patches, these approaches first examine and improve intermediate information signals collected during agent exploration, including \textbf{Reproduction Test}, \textbf{Regression Test}, \textbf{Edit Location}, \textbf{Execution Context}, and \textbf{API Usage}. Although different studies examine different subsets of these signals, no work has systematically compared their relative effectiveness in improving final task success or measured the remaining space for improvement in each signal. To address this gap, this work quantifies the contributions of both \textit{oracle} and agent-extracted signals to final task success and measures the gap between current agent-found signals and their \textit{oracle} counterparts. We hope our findings could suggest future research emphasis in SWE agents. 
We release our source code\textsuperscript{\ensuremath{\ddagger}} and \textit{oracle}-signal dataset\textsuperscript{\ensuremath{\S}} as a reusable methodology to benefit future signal analysis of SWE agents.

\end{abstract}
\begin{figure*}[t]
  \centering
  \includegraphics[width=0.95\linewidth]{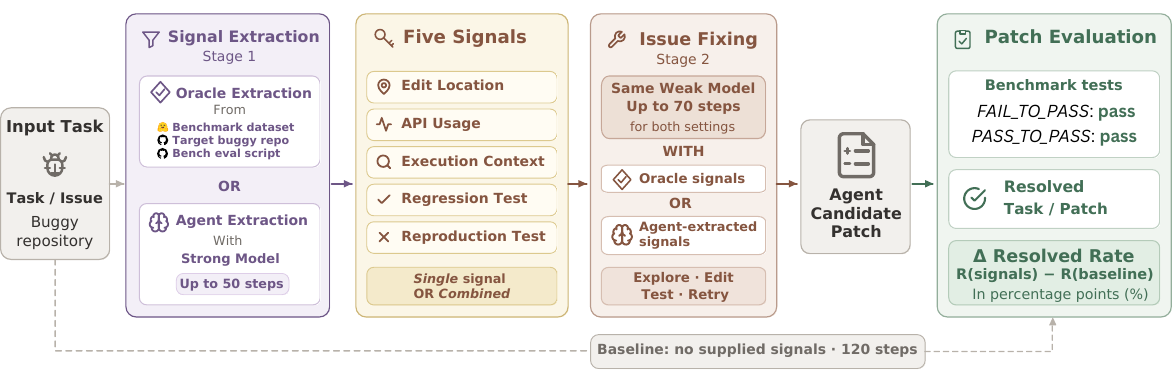} 
  \caption {Overview of the ORACLE-SWE methodology.}
  \label{figure1}
\end{figure*}

\section{Introduction}

Recent code language models (LMs) have progressed from single-file generation~\citep{just2014defects4j,jain2024livecodebench} to agentic, repository-level problem solving, where agents must freely explore and execute an entire codebase. Representative repository-level benchmarks typically derive tasks from native GitHub issues, including the SWE-bench family~\citep{jimenez2023swe, yang2025swebench, yang2025swesmith}, and follow-up works SWE-bench-Live~\citep{zhang2025swe, li2026repolaunch} and Pro~\citep{deng2025swe}, which further increase task diversity and complexity.

To solve repository-level problems, prior work has explored diverse agentic workflows, training methods, and failure analysis. Across these methods, a broad \textbf{consensus} has emerged: improving and evaluating the intermediate information signals acquired during agent exploration is essential, rather than focusing only on the final edit patch. For example, agentic workflows~\citep{yang2024swe, wang2024openhands, xia2024agentless} commonly design separate components of context exploration, target localization, API retrieval, test generation, and test validation before patch generation; training methods~\citep{ma2025sorft, wei2025swe} introduce dedicated loss for code localization and test generation in addition to patch correctness; failure analyses~\citep{meng2024empirical, bouzenia2025understanding} directly assess the quality of these signals. 

These signals are particularly appropriate for SWE tasks, which are knowledge-intensive: the correctness of edit depends heavily on the quality of intermediate information obtained. These SWE-specific signals also provide concrete, quantifiable targets for evaluation and optimization of agentic capabilities, such as reasoning, planning, and memory, which are otherwise difficult to measure directly. Based on a comprehensive literature review (Table~\ref{factor_mapping}), we select five mostly-studied signals: \textbf{Reproduction Test}, \textbf{Regression Test}, \textbf{Edit Location}, \textbf{Execution Context}, and \textbf{API Usage}. Table~\ref{signal-definition} formally defines these signals, each capturing a distinct agentic capability and a distinct type of information required for resolving SWE issues.

Although different studies examine different subsets of these signals, no work has systematically compared their relative effectiveness in improving final task success or quantified the remaining room for improvement in each signal. The answer to this question is important for determining which signal should receive more attention in future research and engineering: without it, researchers may invest substantial effort in developing sophisticated methods for signals with inherently limited potential, while industrial practitioners may waste considerable expensive LLM tokens on extracting and optimizing signals that yield marginal gains. 

In this work, as shown in Figure~\ref{figure1}, we quantify the \textit{ideal} contribution of the \textit{oracle} signals to estimate the effectiveness and room left of improving each signal. Oracle signals are provided to a main agent simulating real-world answers provided by a sub-agent or a stage of an agentic workflow, and the resulting success gain is taken as the contribution. To examine whether the same contribution order holds for realistic agent-extracted signals as agent capabilities improve, we also provide the base agent with signals extracted by a stronger agent. Both settings yield a consistent ordering: Reproduction Test > Execution Context $\sim$ API Usage > Edit Location > Regression Test. Finally, we quantify the gap between agent-extracted signals and their oracle counterparts, revealing how much room remains for improving each capability in current agents.

The contributions of this work are threefold:
\begin{itemize}[leftmargin=1em]
\item \textbf{Identifying the significance of information signals in SWE issue resolution} through comprehensive literature review, and proposing generalizable methods for extracting the \textit{oracle} versions of the five mostly-studied signals to benefit future dataset analysis, supplied with the \textit{source code} reusable for other SWE benchmarks.

\item \textbf{Quantifying the contribution order of the five signals} to suggest future improvement emphasis.

\item \textbf{Providing the oracle-signal \textit{dataset} and a method to quantify the gap between agent-found signals and their oracle counterparts}, further suggesting future improvement directions.

\end{itemize}

In the remainder of this paper, Section~\ref{section2} explains the motivation of selecting the five signals for study, Section~\ref{section3} describes the extraction of each oracle signal, Section~\ref{sec:experiments} introduces the method to quantify contributions of oracle and agent-extracted signals, and Section~\ref{section5} presents the resulting contribution order and the gap between current LM agents and the ultimate oracles.

\begin{figure*}[t]
  \centering
  \includegraphics[width=1.0\linewidth]{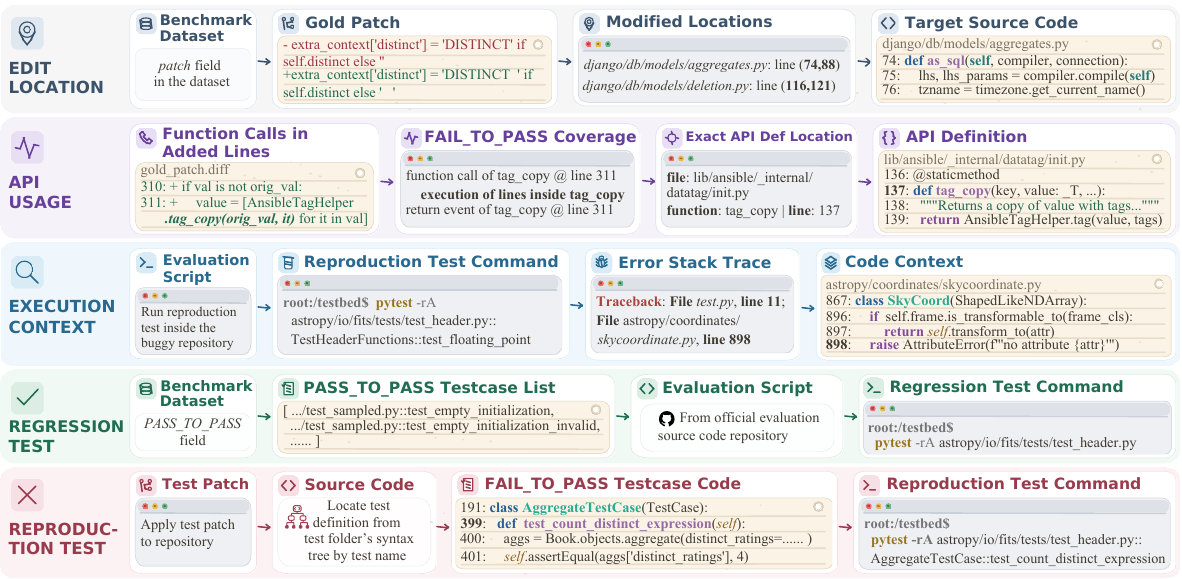} 
  \caption {The extraction method for the oracle versions of the five mostly-studied signals in prior SWE agent works.}
  \label{figure2}
\end{figure*}

\section{Related Works}
\label{section2}

This section reviews which signals each prior work study and based on it summarizes the five most frequently mentioned signals. We therefore select these five mostly-discussed signals for further evaluation.

\subsection{Agentic Workflows for SWE Tasks}
The agentic workflows to resolve SWE tasks structure repository exploration into separate stages or sub-tasks, each obtaining one specific information signal.
Agentless~\citep{xia2024agentless} adopts a canonical staged procedure of code localization, edits, reproduction test generation, and final reproduction and regression test validation.
Free-exploration agents like SWE-agent~\citep{yang2024swe} and OpenHands~\citep{wang2024openhands} follow sequential stages of localization, reproduction, edit and test validation, with a minimal tool set of shell commands and file editor.
Later free agents such as ClaudeCode~\citep{claudecode} and SWE-Edit~\citep{zhang2026sweedit} follow similar multi-stage exploration with a new interface to delegate sub-tasks such as localization, reproduction test generation and patch generation to dedicated sub-agents.
Empirical analysis by~\citep{bouzenia2025understanding} further shows that across different agents and benchmarks, agent trajectories exhibit three recurring stages: \textit{Localization} (explore, search, and identify relevant code regions), \textit{Edit} (reason about the bug and produce a patch), and \textit{Test Execution} (verify edit correctness with reproduction and regression tests) (Figure~\ref{fig:failure} (a)).

Some workflows explicitly decompose SWE tasks into sub-tasks, with each delegated to a sub-agent to obtain some specific information signals. TDFlow~\citep{han2025tdflow} distributes a task across sub-agents dedicated to localization, patch generation, format correction, reproduction-test synthesis, and test-based validation. Several multi-agent approaches further introduce a specialized sub-agent for code-graph construction to enhance contextual reasoning and API retrieval. MarsCode~\citep{liu2024marscode} and Lingma Agent~\citep{ma2025alibaba} build code graphs that combine file trees, abstract syntax trees, and inter-function call relations. GraphLocator~\citep{liu2025graphlocator} augments these graphs with natural-language descriptions to facilitate navigation. AutoCodeRecover~\citep{zhang2024autocoderover} and PatchPilot~\citep{li2025patchpilot} obtain code graphs of function call relations by executing reproduction tests and collecting execution traces.

\subsection{Training Methods for SWE Tasks}

Another direction focuses on directly training LMs to improve the extraction of key intermediate information signals in the SWE workflow, including localization, reproduction-test generation, and patch synthesis.
For instance, SoRFT~\citep{ma2025sorft}, SWE-RL~\citep{wei2025swe} and SWE-Swiss~\citep{he2025swe_swiss} train LMs for single-turn localization, reproduction test generation and patch synthesis with supervised fine-tuning (SFT) followed by reinforcement learning (RL).

More recent efforts such as~\cite{zainullina2025guided, golubev2025training, cao2025skyrl} directly roll out multi-turn agents following the standard order of localization, reproduction, edit, and validation.
The whole rollout trajectories are then collected for rejection-sampling-based SFT and then RL. 
SkyRL ~\citep{cao2025skyrl} extends this framework by incorporating the code graph to enhance contextual reasoning over the repository.

\subsection{Failure Analysis on SWE Tasks}

The failure analysis works also attributes the reason of incorrect patch generation to the failure of attaining these intermediate information. 
SWE-agent~\citep{yang2024swe} categorizes unsuccessful runs into failures in reproduction (failed / overly-specific reproduction), localization, and code generation (incorrect implementations and iterative editing errors) (Figure~\ref{fig:failure} (b)). 
\citep{meng2024empirical} further demonstrate that inaccurate reproduction of the issue can significantly reduce overall task success. 
\cite{chen2025unveiling} analyzes error traces of agent solutions and identify three major categories of errors: (1) code-generation issues such as syntax mistakes, (2) insufficient familiarity with repository context, including import failures or incorrect variable usage, and (3) misuse of external APIs. PAGENT~\citep{xue2025pagent} studies agents' patches and evaluation logs and finds recurring problems including context misunderstanding (e.g., incorrect type transformations), inadequate coverage of corner cases in reproduction, and incorrect usage of internal or external APIs.

These prior studies suggest the critical intermediate information in SWE task resolution can be classified into five signals from two sources:
\begin{enumerate}[leftmargin=1em]
    \item \textbf{From Code Exploration}: Edit Location, relevant Execution Context, and API Usage.

    \item \textbf{From Test Execution Feedback}: Reproduction Test to evaluate requirement implementation and Regression Test to protect original features.
\end{enumerate}

Table~\ref{factor_mapping} summarizes the signals studied in each previous work, while Table~\ref{signal-definition} formally defines these signals, specifying the specific agentic capacity required to obtain each signal and the distinct type of information needed to solve SWE problems that each signal contains. 
\section{Extraction of Oracle Information}
\label{section3}
This section defines the five oracle signals, proposes generalized methods to extract each from SWE benchmarks (Figure~\ref{figure2}), and explains how realistic agents could obtain the same form of signals as the oracles so that the oracles are ultimately obtainable. We provide one example of each oracle signal in Appendix \ref{example_signal}. The extraction is rule-based without randomness, and the correctness has been verified manually.

\subsection{Code Search}

\subsubsection{Edit Location}
The oracle \textit{Edit Location} consists of the code regions where modifications occur in the gold patch.

Each extracted region includes the file path, class / function name and line range, simulating the edit locations found by a localization sub-agent.

\subsubsection{Execution Context}
\textit{Execution Context} refers to the code contexts that call or are called by the problematic locations in the source code when reproducing the issue in an SWE task.
Stack traces summarize the relation of function calls leading to an error and therefore provide a compact representation of the code context.
The oracle \textit{Execution Context} is thus derived from the error stack trace produced when executing reproduction tests on the buggy version of the repository.
Each frame in the trace is represented as a tuple of file path, line number, function name, and surrounding source code (15 lines around the called line), simulating the relevant code relations and snippets submitted by a code graph search sub-agent.
Frames originating from test functions are masked to prevent leakage of ground truth tests.

In Python projects, many failures arise from assertion errors triggered inside test functions, which prevents the stack trace from reaching the underlying source code.
To address this limitation, we insert a custom stack-trace collector at the oracle edit locations and record the deepest reachable call stack during execution for instances without error stack into the source code.
For Golang projects, reproduction tests mostly fail during the build stage, which similarly prevents normal stack-trace collection, so contextual information is entirely obtained from the custom stack-trace collector.

\subsubsection{API usage}
The oracle \textit{API Usage} signal contains all function calls modified or newly introduced in the ground-truth patch.
For each call, we record the function name, its arguments, the approximate region where it should be used, and the corresponding function definition whenever the function is not part of the language's built-in library, simulating the API definitions and suggested ways of usage submitted by a code search agent. 

Simple name matching cannot distinguish function definitions with the same name.
For Python repositories, which is not strictly typed, the correct function definition is inferred through execution coverage, where the invoked API definition just follows the site that calls it.
In contrast, Golang's static typing enables us to resolve the definition through type checking, import analysis, and argument signatures.

\subsection{Testcase Verification}
\subsubsection{Reproduction Test}
The oracle \textit{Reproduction Test} information consists of three components: the command required to execute the reproduction tests, the list of reproduction test names that must pass, and the full source code of those test functions.
These tests correspond to the \textit{Fail-to-Pass} cases in the benchmark datasets.
Each reproduction test typically contains multiple parameterized inputs that cover diverse corner cases, making them a highly informative signal for identifying the bug.

Test patch of a task instance is intentionally applied to the code base for agents to execute the reproduction tests and iteratively refine its edits based on the resulting error outputs. 
The location of each test function is inferred from its name, which generally encodes hierarchical information such as folder, file, class, and function identifiers.
The function bodies are then extracted using syntax tree parsing.

Most modern testing frameworks allow execution of individual testcases.
We therefore provide commands that run only the relevant reproduction tests.
Instances that do not support selective execution are excluded from our evaluation.

Generally the setting simulates that a test generation agent adds tests to the test folder and provides the main agent with the added test contents and test command.

\subsubsection{Regression Test}

The oracle \textit{Regression Test} information includes the command to execute original regression tests in a repo extracted from the benchmark evaluation code, and the list of tests that must remain passed after the fix, which corresponds to the \textit{Pass-to-Pass} cases in the benchmark datasets.
The signal simulates the test command and original passed tests submitted by a verification agent though command trials in a repo.

\section{Experiments}
\label{sec:experiments}

\subsection{Base Agent}
Our experiments require an agent that can freely explore the repository, execute tests to validate the fix, and retry edits based on the test output so that the injected oracle information can be fully utilized during the resolution process.
We therefore adopt SWE-agent~\citep{yang2024swe}, a minimal free agent equipped with two tool interfaces: bash commands and an editor supporting file viewing, replacement, and creation. The max step limit of SWE-agent is set to 120 to balance performance and cost. 

\subsection{Oracle Ablation Study}
We quantify the ideal contribution of each oracle factor via controlled ablations (Figure~\ref{figure1}). Concretely, we append oracle signals to the initial user prompt, emulating answers submitted by sub-agents~\citep{han2025tdflow} or previous workflow stages~\citep{xia2024agentless} to the main/edit agent in real-world issue-resolution settings. The user prompt (Appendix~\ref{sec:appendixB}) explicitly instructs the agent to fully trust the injected information to reduce rollout randomness at best effort, while any knowledge not injected must be acquired through normal exploration. While alternative signal injection / prompting strategies could lead to subtle differences in measured contributions, budget constraints prevent us from exhaustively re-running the study all over again. 

We first measure the success rate and corresponding LM API cost when injecting each factor individually. We then inject combinations of two or more factors to study interactions among oracle signals and to approximate the attainable upper bound when all five factors are perfectly known.

\begin{figure*}[t]
  \centering
  \includegraphics[width=0.95\linewidth]{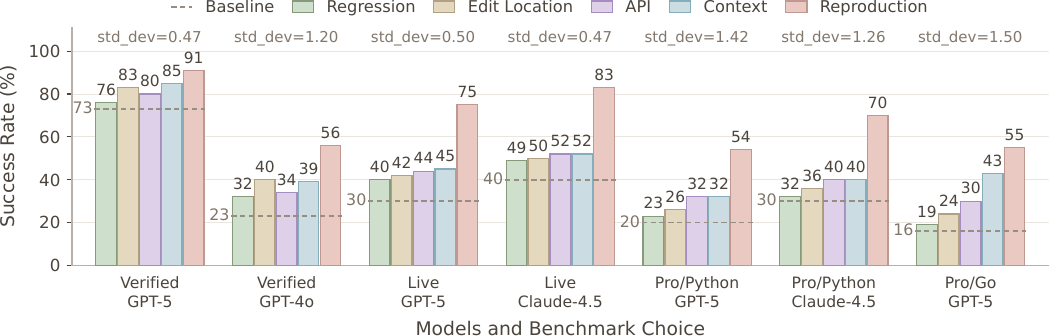} 
  \caption {Ablation study measuring the upper-bound contribution of each oracle factor. Each factor ablation and baseline were run 3 times to reduce randomness. Average success rate of each 3-run and standard deviation of 3 runs averaged across 1 baseline and 5 ablations under a bench-model combination are reported.}
  \label{figure3}
\end{figure*}

\subsection{Validation Study}
Admittedly, it remains uncertain whether the ultimately \textit{correct} signals extracted by real agents though agentic capability improvement can find the ground truths and infer the evaluation criteria of reproduction tests which oracle signals provide. Therefore, in order to validate whether agent-extracted signals when agentic capabilities improve have the same contribution order as the oracle ones and estimate the difficulty of obtaining these signals under realistic conditions, we design a two-stage validation experiment. 

In Stage 1, an agent is prompted to extract each factor using the \textbf{same definition and format} as the oracle signal (Appendix~\ref{val_prompt}); in Stage 2, the agent attempts to solve the task using the factors it extracted. We use a stronger model for Stage 1 to emulate high-quality signal when agent capabilities improve, and a weaker model for Stage 2 (Figure~\ref{figure1}).

As baselines, we evaluate single-agent rollout using the strong or weak model alone. The two-stage pipeline allocates 50 steps to factor extraction and 70 to issue resolution, matching the 120-step budget of the single-agent baseline. For cost control, 100 instances per dataset were sampled for experiment. 

\subsection{Benchmarks and Models}
We evaluate on three datasets: SWE-bench-Verified~\citep{jimenez2023swe}, SWE-bench-Live~\citep{zhang2025swe, li2026repolaunch}, and SWE-bench-Pro~\citep{deng2025swe}. Latest models such as GPT-5 already saturate SWE-bench-Verified with over 70\% success. We therefore also include the more challenging SWE-bench-Live and SWE-bench-Pro datasets released in 2025.

For SWE-bench-Live, we use the verified Python subset for which task feasibility has been confirmed. For SWE-bench-Pro, we evaluate both the Python and Golang subsets to assess cross-language generalization. We filter out instances whose ground-truth patches fail or whose reproduction tests cannot be selectively executed. After filtering, the final datasets contain 459 instances for SWE-bench-Verified, 353 for SWE-bench-Live, 220 for SWE-bench-Pro/Python, and 211 for SWE-bench-Pro/Golang.

We evaluate multiple models to ensure the findings are generalizable. GPT-5-Thinking-Medium and GPT-4o are tested on SWE-bench-Verified. The older GPT-4o exhibits less contamination on the older SWE-bench-Verified benchmark~\citep{zhang2025swe, deng2025swe}, making the results more reliable, and its lower baseline better highlights the absolute impact of each factor. For SWE-bench-Live and Pro, we evaluate both GPT-5 and Claude-4.5-Sonnet. For validation experiment, GPT-4o is the weaker model while GPT-5 is the stronger model for SWE-bench-Verified; for the other benchmarks, GPT-5 is the weaker one while Claude-4.6-Sonnet is the stronger one.
\begin{figure*}[t]
  \centering
  \includegraphics[width=0.90\linewidth]{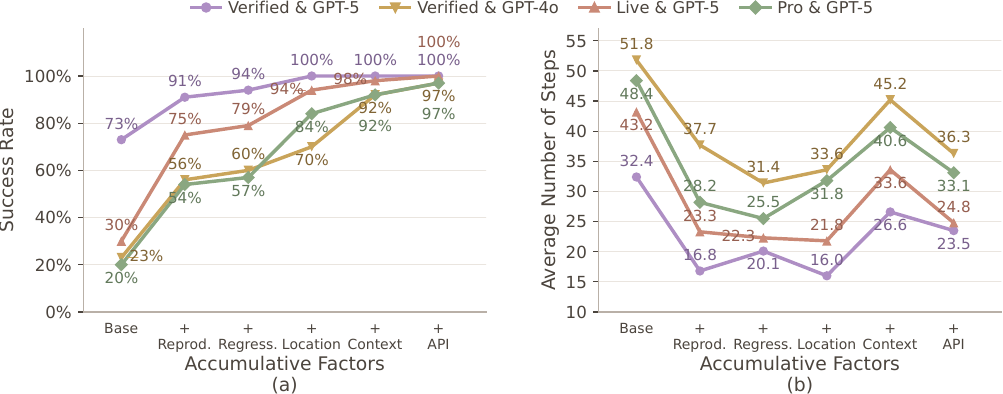} 
  \caption {Accumulative contribution and average number of rollout steps when factors accumulate. Each factor on the x-axis incorporates all factors to its left. Different signal injection orders are tested with no significant performance difference (<2\%).}
  \label{figure4}
\end{figure*}

\section{Results}
\label{section5}

\subsection{Ablation Study on the Upper Bound Contribution of each Oracle Factor}
Figure~\ref{figure3} reports the success rates across the benchmarks and models under study when a single oracle information is provided. The result for execution context is the overall result averaged on native error stack trace and custom stack trace collector (for individual contribution of native stack and custom stack, please refer to Table \ref{context_detail}).
Compared to the baseline where no information is provided, injecting any oracle signal consistently improves the success rate, providing preliminary evidence that the extracted factors are meaningful.

Across models and datasets, the results are highly consistent. On SWE-bench-Live and SWE-bench-Pro, the five factors follow the same ordering: \textbf{Reproduction Test $>$ Execution Context $\sim$ API Usage $>$ Edit Location $>$ Regression Test}. On SWE-bench-Verified, the ordering is unchanged except that Edit Location contributes more than API Usage and Execution Context. Appendix~\ref{task-type} further shows that across the three task categories of SWE tasks evaluated -- bug-fix, feature-implementation and refactoring, the contribution order of the five oracle factors remains consistent.

\textbf{Different Patterns between Error Trace and Custom Trace.} Table \ref{context_detail} shows a clear gap between the contributions of the native error stack and the custom stack. 
The native error stack is much more effective, likely because exceptions raised within the source code are typically explicit failures (e.g., \textit{list index out of range}, \textit{attribute does not exist}). It also provides rich information about the bug-affected execution path and the exact failure point, which substantially improves success rates.

By contrast, custom stacks are used only when native error stacks are unavailable. In these cases, errors are typically implicit (e.g., incorrect calculation result) and rooted in program logic, making the call stack less useful for fixing the issue. Such stack trace is more like a way to locate the edit locations, as suggested by the similar contribution with edit location. In some model-benchmark combinations in Table \ref{context_detail}, the custom stack even contributes slightly less than Edit Location, possibly because such custom stack adds little new information while requiring extra effort to locate the exact edit location. Overall, the results suggest that native error stack traces are highly informative but often unavailable.

\textbf{Indications from API usage.}
We analyze the distribution of API calls extracted from the oracle signals.
Excluding built-in functions (for which we do not provide definitions), internal APIs within the repository account for approximately 82\% of calls, while external APIs from third-party libraries account for about 18\%, suggesting that effective repository-level search for internal utilities is substantially more important than retrieving external documentation. The less contribution of API Utilities on SWE-bench-Verified are because the majority of its APIs originate from built-in libraries and widely-used libraries such as \texttt{django}, \texttt{sympy}, and \texttt{matplotlib}. Modern LMs already possess strong prior knowledge of these APIs, reducing the marginal benefit of providing their API definitions.

\begin{figure*}
    \centering
    \includegraphics[width=0.75\linewidth]{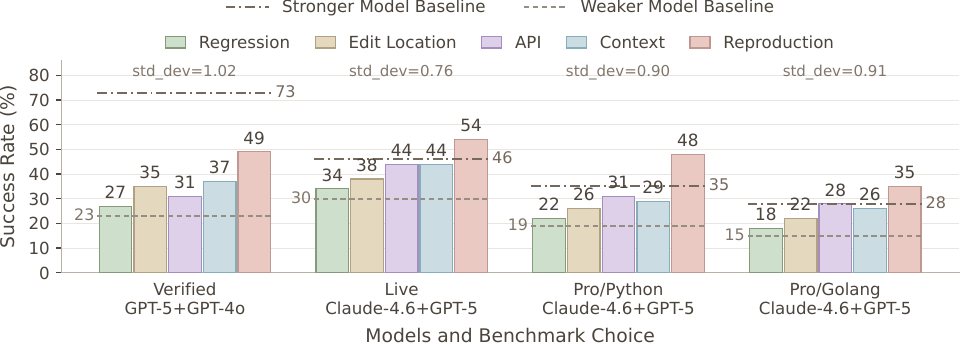}
    \caption{
    \label{figure_5}
    Visualized result of the two-stage validation compared with the single-agent baseline. Labels are formatted as \textit{(strong model) + (weak model)}. Stage 2 and the weak baseline were run 3 times with average success and standard deviation of 3 runs averaged across the 5 factors and the weak baseline reported.}
\end{figure*}

\textbf{The high success rate of reproduction.}
The high contribution of oracle Reproduction Test lies in the precise task requirements it defines. Previous studies~\citep{Chowdhury2024introducing, xia2024agentless, openai2026swebench, openai2026swepro} suggest that GitHub-issue-based SWE benchmarks usually suffer from vague problem statement and overly strict test criteria not obviously required in the problem statement. Resolving a GitHub issue thus requires LM agent to infer strict test criteria from vague natural-language description. Oracle Reproduction Test provides the precise task requirement which is the core information that SWE agents need.

The dominant contribution of reproduction tests raises the concern of whether agents merely satisfy the test criteria rather than solve the underlying issue. To investigate this, we manually inspected 50 agent-predicted patches and prompted GPT-5(High) to label the left. The result shows that \textbf{none} of the prediction patches were considered to cheat. 

We also investigated whether the dominant contribution is because ground truth reproduction tests leak test criteria that LM agent cannot infer in real-world issue resolution. Agent-written reproduction tests in Section~\ref{validation-section} achieve consistently dominant contribution with no ground truth test provided, which shows the ground truth test criteria can genuinely be inferred by agents given strong enough agentic capabilities. We further substitute the ground truth test body with LM-summarized natural language requirements of features to be satisfied with no source code and test criteria like assertions inside. The resulting contribution only drops by 4\%, suggesting the high contribution is due to the clarified task requirements, not the leakage of test criteria that agents cannot infer.

\textbf{Cross-agent generalizability.} ClaudeCode, a more advanced agent with planning and context management tools, is also tested to ensure the generalizability of our conclusion. Under the same setting on SWE-bench-Live, the contribution difference is less than ±2\% for any signal compared to SWE-agent.

\textbf{Trajectory Analysis.}
Appendix~\ref{trajectory-analysis-appendix} presents an analysis of agent rollout trajectories to further explain the impact of oracle factors on agent behavior.

\textbf{Cost Analysis.}
The costs of signal extraction and issue resolution with each signal reflect the practical feasibility to use a signal. Figure~\ref{single_signal_step} reports average step of agent rollout per instance with each signal, Figure~\ref{single_signal_cost} reports LLM API cost breakdowns when running the agent with each signal, while Table~\ref{latency} reports the time latency each oracle-signal extraction takes.

\subsection{Contribution of Factor Combinations}
We further evaluate how the five factors interact when combined.

\textbf{Contribution when factors accumulate.} Figure~\ref{figure4} (a) reports success rates as oracle signals are added incrementally, where each factor on the x-axis includes all factors to its left. With all five factors, all four model-bench pairings reach at least 97\% success, supporting the completeness of our factor set in resolving SWE benchmarks. The small differences across benchmarks in marginal contributions and ordering in Figure~\ref{figure3} and~\ref{figure4} also reflect variation of different benchmarks in task difficulty and distribution, showing that our factor-isolation method can surface dataset-level differences.

Figure~\ref{figure4} (b) shows the average rollout steps under factor accumulation. Steps generally decrease as more signals are added, indicating reduced exploration, but rise again when many factors are combined, suggesting extra iterations are needed to interpret and integrate multiple signals. Average cost is reported in Table~\ref{accumulative_cost} in Appendix. Cost decreases less than steps because oracle signals increase context length; for example, test execution output can substantially lengthen LM inputs. An alternative accumulation order in Figure~\ref{another_direction} shows monotonic success rate increase and consistent step count and cost pattern.

\textbf{Contribution of two-factor combination.} Since the accumulation experiment suggests that combined-factor contributions are not additive, pairwise combinations are further evaluated in Table~\ref{two_factors}. Results suggest that improving multiple fac together yields substantially larger gains than any single factor alone, because the outcome of one signal may help the model understand another signal. For example, when combining Edit Location with test signals, test outcomes help the model infer how to modify the identified locations. 

\begin{table*}[t]

  \renewcommand{\arraystretch}{1.18}
  \setlength{\tabcolsep}{8pt}

  \begin{tabularx}{\textwidth}{
      @{}
      >{\raggedright\arraybackslash}p{6.1cm}
      >{\raggedleft\arraybackslash}p{1.6cm}
      >{\raggedright\arraybackslash}p{5cm}
      >{\raggedleft\arraybackslash}p{1.6cm}
      @{}
    }
    \toprule
    \textbf{Metric} & \textbf{Accuracy}
      & \textbf{Metric} & \textbf{Accuracy} \\
    \midrule

    \rowcolor{gray!12}[0pt][\tabcolsep]
    \multicolumn{4}{@{}l}{\textbf{Reproduction Test\#}} \\
    Fully reproduce
      & 20.5\%
      & Fail to cover some cases
      & 55.5\% \\
      Fail to reproduce the issue
      & 12.5\%
      &
      Do not know how to run the repo
      & 11.5\% \\
    \addlinespace[3pt]

    \rowcolor{gray!12}[0pt][\tabcolsep]
    \multicolumn{4}{@{}l}{\textbf{Regression Test\#}} \\
    Run all related tests correctly
      & 52.0\% &
    Miss some related tests
      & 17.5\%  \\
    Miss all related tests
      & 17.0\% &
    Do not know how to run the repo
      & 13.5\% \\
    \addlinespace[3pt]

    \rowcolor{gray!12}[0pt][\tabcolsep]
    \multicolumn{4}{@{}l}{\textbf{Edit Location}} \\
    File-path precision
      & 69.0\%
      & File-path recall
      & 74.9\% \\
    Line-number precision
      & 43.2\%
      & Line-number recall
      & 48.2\% \\
    \addlinespace[3pt]

    \rowcolor{gray!12}[0pt][\tabcolsep]
    \multicolumn{4}{@{}l}{\textbf{Execution Context}} \\
    File-path precision*
      & 38.2\%
      & File-path recall*
      & 72.7\% \\
    Function precision**
      & 26.4\%
      & Function recall**
      & 65.1\% \\
    \addlinespace[3pt]

    \rowcolor{gray!12}[0pt][\tabcolsep]
    \multicolumn{4}{@{}l}{\textbf{API Usage}} \\
    API function-name precision
      & 41.8\%
      & API function-name recall
      & 41.9\% \\
    API definition localization accuracy\$
      & 96.5\%
      & & \\
    \bottomrule
  \end{tabularx}
    \caption{\label{validation-accuracy}
    Accuracy of the signals extracted by the agent compared to the oracle signals. 
    \# The accuracy of Reproduction Test and Regression Test requires semantic understanding, which we first labeled using an LLM (GPT-5-Thinking-High) then checked manually.
    * including all files in the execution stack;
    ** including all functions in the execution stack;
    \$ Accuracy of API function definition localization is calculated among the agent-extracted API functions with correct function names.}
\end{table*}

\subsection{Validation Experiment}
\label{validation-section}
\textbf{Consistent order with oracle version.} Figure~\ref{figure_5} shows the contribution of agent-extracted signals over single-agent baselines. Using the same weak model in both stages improves success by at most 4\% over the weak baseline, indicating that most gains come from improved agentic ability brought by the strong model. The ordering aligns with oracle patterns, suggesting oracle contributions can indicate the contribution order of agent-extracted signals as agentic capabilities improve. Execution Context contributes slightly less than its oracle counterpart, likely because it is harder to extract in practice, especially when error traces are unavailable as discussed above.

Surprisingly, the two-stage pipeline exceeds the \textbf{stronger} model baseline with agent-extracted reproduction, highlighting the algorithmic potential of this combined strategy.

\textbf{Cost Analysis.} 
As shown in Table~\ref{validation_cost}, when using agent to extract the signals, Reproduction Test is cheapest for SWE-bench-Verified, and Edit Location for others. For issue resolution with extracted information, Edit Location is cheapest on SWE-bench-Verified, whereas Reproduction Test is cheapest on the others.

\textbf{Accuracy of agent-extracted signals} compared to oracle signals is presented in Table~\ref{validation-accuracy}. The accuracies explain the contribution gap between oracle signals and agent-extracted signals -- agent misses some cases or even does not know how to run the repos when extracting test signals; and undershoots relevant code locations when extracting Edit Location, Execution Context and API Utilities. 
\section{Conclusion}
This work quantifies the relative effectiveness and the room left of improving the five mostly-studied information signals in SWE task resolution to suggest the emphasis of future improvements. We further provide the oracle-signal datasets and the method to quantify the gap between agent-extracted signals and oracle signals to facilitate future improvements in SWE agents.
\section*{Limitations}
This work selects five intermediate information signals which are most frequently discussed by prior SWE agent works for study.
The categorization is partly subjective and intended to synthesize and extend previous observations, rather than to represent a definitive or exhaustive decomposition of the knowledge or a full coverage of agentic capabilities required for SWE issue resolution.

Admittedly, the oracle signal study is based on the assumption that agent can ultimately get the same form of correct answers when corresponding agentic capabilities continue to approve, although the consistent contribution order of agent extracted signals supports this assumption.

\clearpage
\bibliography{custom}

\appendix

\section{Supplementary Information for Related Works}
\subsection{Analysis of proportions of related factors in related works}
\label{factor_in_related_works}

\begin{figure*}[t]
    \centering
    \includegraphics[width=\linewidth]{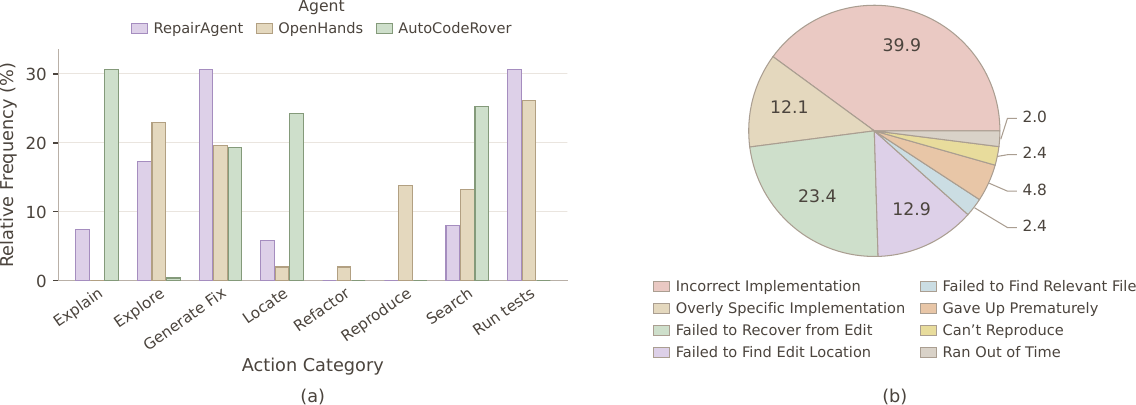}
    \caption{(a) Proportion of SWE agents' exploration pattern, referred from~\citep{bouzenia2025understanding}. (b) Proportion of SWE-Agent's failure modes, referred from~\citep{yang2024swe}.}
    \label{fig:failure}
\end{figure*}

Please refer to Figure ~\ref{fig:failure}.

\subsection{Mapping of Related Works to Corresponding Factors}

Please refer to Table \ref{factor_mapping}.

\section{Problem Formulation}

Table~\ref{factor_mapping} summarizes the information signals/factors on which each related work on SWE agents focuses. We then formally define the five information signals essential for SWE task resolution in Table~\ref{signal-definition}, specifying the aspect of necessary information for solving SWE issues that each signal contains, and the agentic capability required to obtain each signal.

\section{Agent Trajectory Analysis when a Single Factor is Injected}

\label{trajectory-analysis-appendix}

Trajectory level evidence of the impact of each oracle factor on agent behavior is shown in Table~\ref{trajectory-analysis-table}. In the trajectories, the average numbers of key agent actions per instance, including average number of read actions, number of files read, number of edit actions, number of test command execution actions, were counted to track the agent behavior changes. Edit Location and Reproduction Test significantly reduce the search space (reduced number of files read) and provide more meaningful test validation feedback (reduced number of test commands). Execution Context does not reduce the search space because it instructs the agent to read more relevant files around the edit locations; however, it greatly reduces the number of edit operations due to its contribution to contextual reasoning and understanding. Regression does not affect the agent actions much compared to the baseline as it is only a safety guarantee. API utilities increase the number of read and edit operations because it requires the agent to read additional files containing the API utilities and figure out how to use the specified utility functions through multiple trials.

\section{Contribution of Oracle Signals by Task Type}
\label{task-type}
The SWE tasks from the three benchmarks used for evaluation can be classified into three categories -- "bug-fix", "feature-implementation" and "refactoring", as shown in Table~\ref{task_type_proportion}. The classification method follows~\cite{rashid2025swe}, which is first labeled by a reasoning LLM (GPT-5-Thinking-High) and then manually checked. 
Table~\ref{contribution_by_type} shows the contribution of the five oracle signals by task type under the three benchmarks evaluated. It can be seen that feature implementation tasks are more difficult. However, despite the task difficulty difference, i.e., the different success rate baselines, the contribution order of the five signals by task type remains consistent with the overall contribution order, supporting the generalizability of our findings.

\section{Extended Result on Execution Context}

The contribution of native error stack trace and custom stack collector into edit locations result in different contributions, see Table~\ref{context_detail}.

\section{Cost of Agent Rollout with each Signal and Cost of Signal Extraction}
\label{sec:appendixD}

The cost of agent rollout when supplying the agent with each signal and also the cost (here the execution time) of extracting each signal reveal the practical difficulty and feasibility of using each signal.

\begin{figure*}[t]
  \centering
  \includegraphics[width=1.0\linewidth]{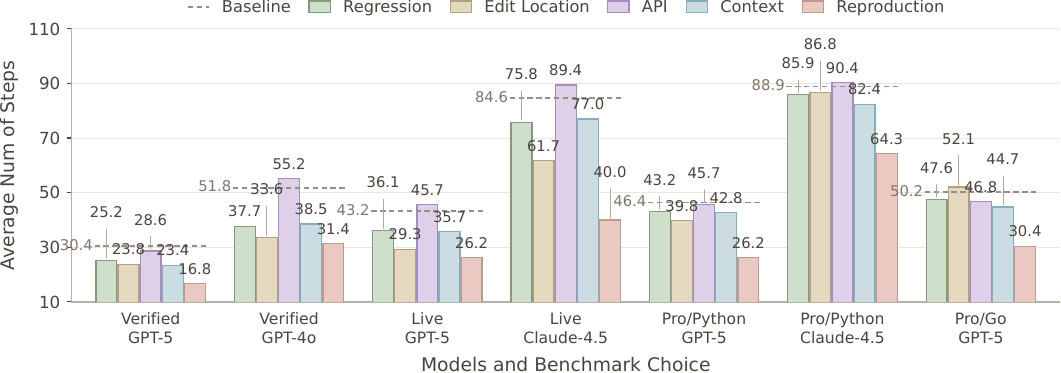} 
  \caption {\label{single_signal_step}
  Average number of steps for single factors.
  }
  \label{figure6}
\end{figure*}

\begin{figure*}[t]
  \centering
  \includegraphics[width=1.0\linewidth]{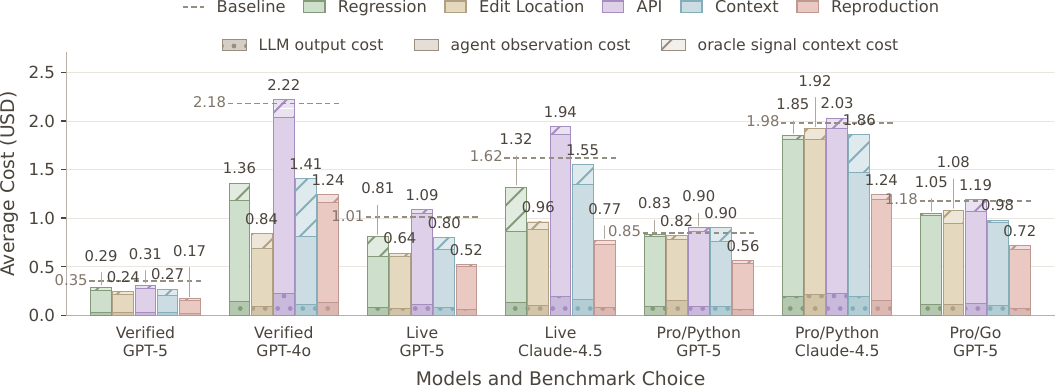} 
  \caption {\label{single_signal_cost}
  Average cost of rollout for single factors. In each bar, from top to bottom, the cost is divided into three parts: the input cost of oracle signals; the cost of subsequent agent observations; the cost of LLM output.
  }
  \label{figure5}
\end{figure*}

Figure \ref{figure5} shows the average cost of solving each instance when one factor is provided.
Figure \ref{figure6} shows the average number of steps of solving each instance when one factor is provided. Cost of API is higher than others because the agent needs multiple rounds of edits and test validation to use API utilities correctly from our observation. The slightly higher cost for Execution Context than for Edit Location suggests the agent spends additional steps exploring the repository to locate the exact edit locations and understand the provided function-call relationships. Regression has higher cost because the requirement of running regression tests adds additional steps but contributes little to the actual issue resolution, and the lengthy output of running regression tests such as long list of test statuses would lead to increase in LM input and thus increase in cost.

Table~\ref{latency} reports the time latency each signal extraction takes, with breakdowns on the modules that take the longest time when extracting each signal. Experiments were conducted on one Linux machine with 64 CPUs and 500GB RAM. 

\section{Extended Results of Combined Factors.}
\label{sec:appendixE}

Figure~\ref{figure4} reports the average number of steps as factors accumulate, while Table~\ref{accumulative_cost} presents the corresponding average cost.
Each successive factor on the right includes all factors to its left.

Since Figure~\ref{figure4} suggests that the contribution of combined factors is not additive, we further evaluate pairwise combinations in Table~\ref{two_factors}.
The results show that combining factors such as Reproduction Test, Edit Location, and API yields substantially larger improvements than any single factor alone, indicating that jointly improving multiple factors is more effective than optimizing them in isolation.
In contrast, Regression Test continues to provide limited benefit even when combined with other factors.

\begin{figure*}[t]
  \includegraphics[width=0.95\linewidth]{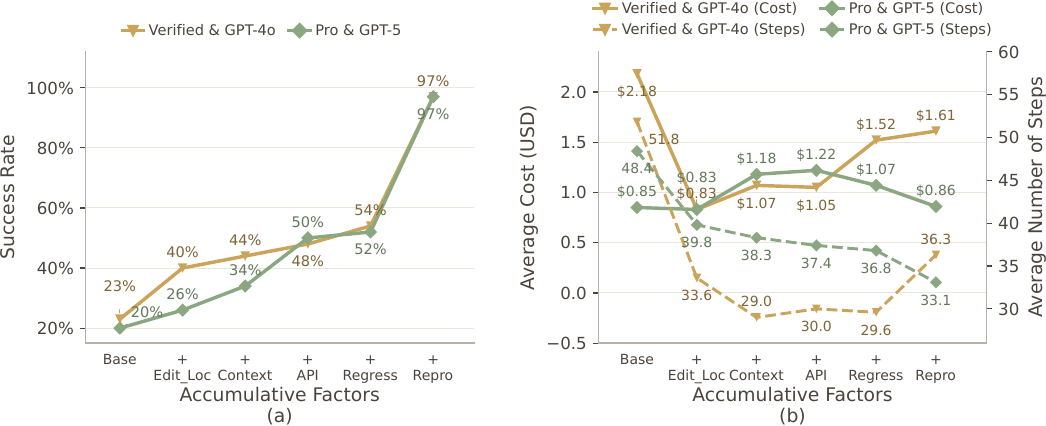}
  \caption {Accumulative contribution and average cost with factors accumulated from another direction.}
  \label{another_direction}
\end{figure*}

To further examine this, we also evaluate an alternative accumulation order (Figure~\ref{another_direction}). The success rate consistently increases as additional factors are introduced, demonstrating the positive effect of combining multiple information sources.

It is important to note that average cost does not always decrease when multiple factors are combined.
Providing additional oracle signals increases context length and may require broader exploration to utilize them effectively.
For example, test execution logs can be lengthy and significantly expand the prompt.
Similarly, the average number of steps does not necessarily decrease, as the agent may require additional steps to understand and integrate multiple factors correctly.

\section{Validation Experiment Result Details}
\label{sec:appendixF}
For the success rate gain of agent-extracted signals and cost of the two stages, please refer to Table~\ref{validation_cost}. For the accuracy of agent-extracted signals compared to oracle signals, please refer to Table~\ref{validation-accuracy}. The accuracy of Reproduction Test and Regression Test requires semantic understanding, which we first labeled using a reasoning model (GPT-5-Thinking-High) and then checked manually.

\begin{table*}
  \centering
  \begin{tabularx}{1.0\textwidth}{p{0.38\textwidth}p{0.10\textwidth}p{0.07\textwidth}p{0.06\textwidth}p{0.08\textwidth}p{0.09\textwidth}p{0.04\textwidth}}
    \hline
    \textbf{Related Work} & \textbf{Type} & \textbf{Repro-duction} & \textbf{Regre-ssion} & \textbf{Location} & \textbf{Execution Context} & \textbf{API} \\
    \hline
    Agentless \citep{xia2024agentless} & Workflow & $\checkmark$ & $\checkmark$ & $\checkmark$ & $\times$ & $\times$ \\
    SWE-agent\\\citep{yang2024swe} & Workflow & $\checkmark$ & $\times$ & $\checkmark$ & $\bigcirc$ & $\times$ \\
    OpenHands \citep{wang2024openhands} & Workflow & $\checkmark$ & $\checkmark$ & $\checkmark$ & $\bigcirc$ & $\times$ \\
    Claude Code \citep{claudecode} & Workflow & $\checkmark$ & $\bigcirc$ & $\checkmark$ & $\bigcirc$ & $\bigcirc$ \\
    TDFlow \citep{han2025tdflow} & Workflow & $\checkmark$ & $\times$ & $\checkmark$ & $\times$ & $\times$ \\
    MarsCode \citep{liu2024marscode} & Workflow & $\times$ & $\times$ & $\checkmark$ & $\checkmark$ & $\checkmark$ \\
    Lingma Agent \citep{ma2025alibaba} & Workflow & $\times$ & $\times$ & $\checkmark$ & $\checkmark$ & $\checkmark$ \\
    GraphLocator \citep{liu2025graphlocator} & Workflow & $\times$ & $\times$ & $\checkmark$ & $\checkmark$ & $\times$ \\
    AutoCodeRecover \citep{zhang2024autocoderover} & Workflow & $\checkmark$ & $\times$ & $\checkmark$ & $\checkmark$ & $\times$ \\
    PatchPilot \citep{li2025patchpilot} & Workflow & $\checkmark$ & $\times$ & $\checkmark$ & $\checkmark$ & $\times$ \\
    SoRFT \citep{ma2025sorft} & Training & $\times$ & $\times$ & $\checkmark$ & $\times$ & $\times$ \\
    SWE-RL \citep{wei2025swe} & Training & $\times$ & $\times$ & $\checkmark$ & $\times$ & $\times$ \\
    SWE-Swiss \citep{he2025swe_swiss}  & Training & $\checkmark$ & $\times$ & $\checkmark$ & $\times$ & $\times$ \\
    \citep{zainullina2025guided}  & Training & $\checkmark$ & $\times$ & $\checkmark$ & $\bigcirc$ & $\times$ \\
    \citep{golubev2025training}  & Training & $\checkmark$ & $\times$ & $\checkmark$ & $\bigcirc$ & $\times$ \\
    SkyRL \citep{cao2025skyrl}   & Training & $\checkmark$ & $\times$ & $\checkmark$ & $\checkmark$ & $\times$ \\
    \citep{bouzenia2025understanding} & Analysis & $\checkmark$ & $\checkmark$ & $\checkmark$ & $\bigcirc$ & $\times$ \\
    \citep{meng2024empirical} & Analysis & $\checkmark$ & $\times$ & $\times$ & $\times$ & $\times$ \\
    \citep{chen2025unveiling} & Analysis &  $\times$ & $\times$ & $\times$ & $\checkmark$ & $\checkmark$ \\
    PAGENT \citep{xue2025pagent} & Analysis &  $\times$ & $\times$ & $\times$ & $\checkmark$ & $\checkmark$ \\
    \hline
  \end{tabularx}
  \caption{\label{factor_mapping}
    The signals each SWE agent work focuses on in \textbf{Related Works}. "Workflow" means agentic workflow/harness works; "Training" means LM agentic training works; "Analysis" means failure pattern analysis works.
  }
\end{table*}

\newcolumntype{L}[1]{%
  >{\raggedright\arraybackslash
    \hsize=#1\hsize
    \linewidth=\hsize}X%
}

\begin{table*}[t]
  \centering

  \small
  \setlength{\tabcolsep}{6pt}
  \renewcommand{\arraystretch}{1.35}

  \begin{tabularx}{\textwidth}{
    @{}
    L{0.80}
    L{0.70}
    L{1.15}
    L{1.10}
    L{1.25}
    @{}
  }
    \toprule

    \rowcolor{gray!15}
    \textbf{Information source}
      & \textbf{Signal}
      & \textbf{Definition}
      & \textbf{Information captured}
      & \textbf{Agentic capability required / behavior specs} \\

    \midrule

    Command execution feedback
      & \textbf{Reproduction test}
      & Tests that evaluate whether the new requirements are met.
      & Task requirements
      & Task understanding \\

    \addlinespace[2pt]

    Command execution feedback
      & \textbf{Regression test}
      & Tests that verify existing functionality remains unaffected.
      & Safety guarantees for the edit
      & Safety-aware editing \\

    \midrule

    Source-code exploration feedback
      & \textbf{Edit location}
      & Source-code locations that must be modified.
      & The target of the task
      & Target localization \\

    \addlinespace[2pt]

    Source-code exploration feedback
      & \textbf{Execution context}
      & Code that executes before and after the edit location.
      & Repository components affected by the task
      & Reasoning of code relations \\

    \addlinespace[2pt]

    Source-code exploration feedback
      & \textbf{API Usage}
      & Internal / external functions that should be used in the edit.
      & Knowledge required to perform the edit
      & Knowledge retrieval \\

    \bottomrule
  \end{tabularx}
    \caption{\label{signal-definition}
    The definition of critical information signals in SWE task resolution. Each row specifies the corresponding aspect of necessary information for solving SWE issues that each signal contains (column \textit{Information Captured}), and the agentic capability / behavior specs required to obtain each signal (column \textit{Agentic capability required / behavior specs}).}
\end{table*}

\begin{table*}[t]
\centering
\small
\setlength{\tabcolsep}{3pt}
\renewcommand{\arraystretch}{1.4}

\begin{tabularx}{\textwidth}{
    >{\hsize=1.10\hsize\linewidth=\hsize
      \raggedright\arraybackslash}X
    >{\hsize=0.80\hsize\linewidth=\hsize
      \centering\arraybackslash}X
    >{\hsize=1.05\hsize\linewidth=\hsize
      \centering\arraybackslash}X
    >{\hsize=0.95\hsize\linewidth=\hsize
      \centering\arraybackslash}X
    >{\hsize=0.95\hsize\linewidth=\hsize
      \centering\arraybackslash}X
    >{\hsize=1.05\hsize\linewidth=\hsize
      \centering\arraybackslash}X
}
\hline
\textbf{Dataset} &
\textbf{Regression} &
\textbf{Location} &
\textbf{API} &
\textbf{Context} &
\textbf{Reproduction} \\
\hline

\textbf{SWE-bench-Verified} &
0.01 & 0.3 & 45.3 & 31.1 & 28.3 \\

\textbf{SWE-bench-Live/\newline Verified} &
0.02 & 0.4 & 55.1 & 44.4 & 40.6 \\

\textbf{SWE-bench-Pro/\newline Python} &
0.02 & 0.4 & 98.7 & 89.6 & 80.9 \\

\textbf{SWE-bench-Pro/\newline Go} &
0.02 & 0.4 & 5.8* & 108.7 & 90.2 \\

{\footnotesize\bfseries\itshape
Main bottleneck\newline
(Share of latency)} &

{\footnotesize\itshape
Evaluation script\newline
iteration (98\%)} &

{\footnotesize\itshape
Syntax tree traversal \newline
(92\%); diff patch \newline
parsing (6\%)} &

{\footnotesize\itshape
Run tests (70\%);\newline
capture/write\newline
coverage (27\%)} &

{\footnotesize\itshape
Run tests (80\%);\newline
capture/write\newline
stack trace (16\%)} &

{\footnotesize\itshape
Run test command\newline
to verify\newline
correctness (98\%)} \\

\hline
\end{tabularx}

\caption{
Time latency to extract each oracle signal in \textbf{\textit{minutes}}
for the whole benchmark, and the module that takes the most time
in the extraction of each.
Experiments were conducted on one Linux machine with 64 CPUs
and 500GB RAM.
*GoLang uses static syntax tree analysis to find API definition
instead of coverage, thus taking much less time compared to Python.
}
\label{latency}
\end{table*}

\clearpage

\begin{table*}
  \centering
    \begin{tabular}{lcccc}
    \hline
    Dataset & Dataset size & Bug-fix & New-feature & Refactoring \\
    \hline
    SWE-bench-Verified & 459 & 75.6\% (347/459) & 15.7\% (72/459) & 8.7\% (40/459) \\
    SWE-bench-Live/Verified & 353 & 73.7\% (260/353) & 18.4\% (65/353) & 7.9\% (28/353) \\
    SWE-bench-Pro/Python & 220 & 53.2\% (117/220) & 34.1\% (75/220) & 12.7\% (28/220) \\
    SWE-bench-Pro/Go & 211 & 45.5\% (96/211) & 42.7\% (90/211) & 11.8\% (25/211) \\
    Sum & 1243 & 66.0\%(820/1243) & 24.5\%(305/1243) & 9.7\%(121/1243) \\
    \hline
    \end{tabular}
    \caption{\label{task_type_proportion}
    Proportion of different types of SWE tasks in the benchmarks under evaluation.
  }
\end{table*}

\begin{table*}
  \centering
    \begin{tabular}{lcccccc}
    \hline
    Task type & Baseline & Regression & Location & API & Context & Reproduction \\
    \hline
    Bug fix	& 23.2 & 29.2 (+6.0) & 37.1 (+13.9) & 37.7 (+14.5) & 39.8 (+16.6) & 67.3 (+44.1) \\
    New feature	& 13.3 & 15.6 (+2.3) & 21.6 (+8.3) & 21.2 (+7.9) & 22.6 (+9.3) & 59.5 (+46.2) \\
    Refactor & 28.4 & 29.9 (+1.5) & 39.0 (+10.6) & 50.0 (+21.6) & 40.9 (+12.5) & 70.4 (+42.0) \\
    \hline
    \end{tabular}
    \caption{\label{contribution_by_type}
    The success rate gain of the five oracle signals under each task type. For the three SWE task types, despite the task difficulty difference, the contribution order of the five signals remain consistent.
  }
\end{table*}

\begin{table*}
  \centering
    \begin{tabular}{lcc cc cc c}
    \hline
    & \multicolumn{2}{c}{Verified} & \multicolumn{2}{c}{Live} & \multicolumn{2}{c}{Pro/Python} & Pro/Go \\
    \hline
    & GPT-5 & GPT-4o & GPT-5 & Claude-4.5 & GPT5 & Claude-4.5 & GPT-5 \\
    \textbf{Native Error} \\
    Num. instances & 159 & 159 & 72 & 72 & 50 & 50 & 0 \\
    Resolved (\%) & 87.4 & 42.1 & 64.0 & 66.7 & 48.0 & 56.0 & -- \\
    Avg. Cost (usd) & 0.29 & 1.53 & 0.83 & 1.60 & 0.87 & 1.88 & -- \\
    Avg. Step & 25.5 & 39.3 & 44.3 & 84.2 & 44.0 & 84.0 & -- \\
    \textbf{Custom Collector} \\
    Num. instances & 300 & 300 & 281 & 281 & 170 & 170 & 211 \\
    Resolved (\%) & 84.0 & 38.0 & 39.9 & 48.0 & 27.1 & 35.3 & 45.0 \\
    Avg. Cost (usd) & 0.26 & 1.36 & 0.79 & 1.54 & 0.91 & 1.85 & 1.47 \\
    Avg. Step & 22.3 & 38.0 & 33.5 & 75.9 & 42.3 & 81.7 & 58.3 \\
    \textbf{Averaged} \\
    Resolved (\%) & 85.2 & 39.4 & 45.0 & 51.8 & 31.7 & 40.0 & 45.0 \\
    Avg. Cost (usd) & 0.27 & 1.41 & 0.80 & 1.55 & 0.90 & 1.86 & 1.47 \\
    Avg. Step & 23.4 & 38.5 & 35.7 & 77.0 & 42.8 & 82.4 & 58.3 \\
    \hline
    \end{tabular}
    \caption{\label{context_detail}
    Contribution of execution context for error stack and custom stack respectively.
  }
\end{table*}

\clearpage

\begin{table*}
  \centering
    \begin{tabular}{lcccc}
    \hline
	& \# read actions & \# files read & \# file edit actions & \# test command execution \\
    \hline
    baseline & 14.5 & 8.0 & 8.8 & 11.5 \\
    location & \textbf{11.1} & \textbf{4.0} & 9.3 & \textbf{8.1} \\
    context & 11.8 & 6.8 & \textbf{6.5} & 9.0 \\
    api & \textit{15.0} & 6.5 & \textit{10.2} & 10.1 \\
    regression & 14.3 & 7.7 & 8.3 & 10.7 \\
    reproduction & \textbf{9.3} & \textbf{4.4} & 6.2 & \textbf{4.7} \\
    \hline
    \end{tabular}
    \caption{\label{trajectory-analysis-table}
    Average number of key agent actions per instance, including the average number of file read actions, number of files read, number of file edit actions, and number of test command execution actions. '\#' means 'the average number of' in the table.
  }
\end{table*}

\begin{table*}[t]
  \centering
  \setlength{\tabcolsep}{8pt}
  \renewcommand{\arraystretch}{1.15}

  \begin{tabular}{lcccc}
    \toprule
    \textbf{Signals}
      & \makecell{\textbf{Verified}\\\textbf{GPT-5}}
      & \makecell{\textbf{Verified}\\\textbf{GPT-4o}}
      & \makecell{\textbf{Live}\\\textbf{GPT-5}}
      & \makecell{\textbf{Pro/Python}\\\textbf{GPT-5}} \\
    \midrule

    Base
      & 0.35
      & 2.18
      & 1.11
      & 0.85 \\

    + Reproduction
      & \makecell{0.17\\{\scriptsize (0.02, 0.13, 0.02)}}
      & \makecell{1.24\\{\scriptsize (0.08, 1.03, 0.13)}}
      & \makecell{0.52\\{\scriptsize (0.02, 0.44, 0.06)}}
      & \makecell{0.56\\{\scriptsize (0.03, 0.47, 0.06)}} \\

    + Regression
      & \makecell{0.23\\{\scriptsize (0.05, 0.14, 0.04)}}
      & \makecell{1.77\\{\scriptsize (0.26, 1.40, 0.11)}}
      & \makecell{0.47\\{\scriptsize (0.22, 0.20, 0.05)}}
      & \makecell{1.09\\{\scriptsize (0.05, 0.93, 0.11)}} \\

    + Location
      & \makecell{0.15\\{\scriptsize (0.07, 0.06, 0.02)}}
      & \makecell{1.21\\{\scriptsize (0.40, 0.70, 0.11)}}
      & \makecell{0.83\\{\scriptsize (0.25, 0.49, 0.09)}}
      & \makecell{0.84\\{\scriptsize (0.09, 0.68, 0.07)}} \\

    + API
      & \makecell{0.31\\{\scriptsize (0.09, 0.16, 0.06)}}
      & \makecell{1.98\\{\scriptsize (0.52, 1.32, 0.14)}}
      & \makecell{1.18\\{\scriptsize (0.28, 0.79, 0.11)}}
      & \makecell{0.96\\{\scriptsize (0.11, 0.76, 0.09)}} \\

    + Context
      & \makecell{0.18\\{\scriptsize (0.11, 0.05, 0.02)}}
      & \makecell{1.61\\{\scriptsize (0.81, 0.72, 0.08)}}
      & \makecell{0.50\\{\scriptsize (0.34, 0.11, 0.05)}}
      & \makecell{0.86\\{\scriptsize (0.21, 0.58, 0.07)}} \\

    \bottomrule
  \end{tabular}

  \caption{
    Average LLM API cost in USD for the accumulative contribution experiment in Figure~\ref{figure4}. Each row includes all
    signals listed above it. Parentheses report the three LLM API cost components: (input cost of oracle signal contexts, input cost of agent observations, LLM output cost).
  }
  \label{accumulative_cost}
\end{table*}

\begin{table*}
    \small
    \centering
    \begin{tabular}{lccc ccc}
    \hline
     & \multicolumn{3}{c}{\textbf{Verified\&GPT-4o}} 
     & \multicolumn{3}{c}{\textbf{Pro-Python\&GPT-5-Thinking-Med}} \\
    \hline
     \textbf{Signals} & \textbf{Resolved (\%)} & \textbf{Avg. Cost (\$)} & \textbf{Avg. Step} 
     & \textbf{Resolved (\%)} & \textbf{Avg. Cost (\$)} & \textbf{Avg. Step} \\
    Edit Loc. + Reprod. & 64.9 & \makecell{1.19 \\{\scriptsize (0.26, 1.40, 0.11)}} & 34.0 & 80.0 & \makecell{0.90 \\{\scriptsize (0.07, 0.74, 0.09)}} & 31.8 \\
    Context + Reprod. & 64.9 & \makecell{2.15\\{\scriptsize (0.69, 1.35, 0.11)}} & 40.1 & 79.1 & \makecell{0.78 \\{\scriptsize (0.17, 0.53, 0.08)}} & 22.8  \\
    Reprod. + API & 68.0 & \makecell{1.20\\{\scriptsize (0.32, 0.78, 0.10)}} & 33.8 & 60.0 & \makecell{0.87 \\{\scriptsize (0.07, 0.71, 0.09)}} & 29.1 \\
    Regress. + Reprod. & 58.9 & \makecell{1.77\\{\scriptsize (0.26, 1.28, 0.13)}} & 31.4 & 56.8 & \makecell{1.09\\{\scriptsize (0.05, 0.93, 0.11)}} & 45.9 \\
    Edit Loc. + API & 43.6 &\makecell{0.72\\{\scriptsize (0.34, 0.31, 0.07)}} & 31.3 & 48.2 & \makecell{1.09 \\{\scriptsize (0.08, 0.88, 0.13)}} & 43.1 \\
    Context + API & 43.6 & \makecell{1.60\\{\scriptsize (0.78, 0.66, 0.16)}} & 36.5 & 44.1 & \makecell{1.14 \\{\scriptsize (0.18, 0.84, 0.12)}} & 39.4 \\
    Context + Edit Loc. & 44.0 & \makecell{1.07 \\{\scriptsize (0.75, 0.20, 0.12)}} & 29.0 & 34.1 & \makecell{1.18 \\{\scriptsize (0.18, 0.88, 0.12)}} & 38.3 \\
    Regress. + API & 40.3 & \makecell{1.29 \\{\scriptsize (0.39, 0.77, 0.13)}} & 39.0 & 38.2 & \makecell{1.07 \\{\scriptsize (0.06, 0.90, 0.11)}} & 44.7 \\
    Edit Loc. + Regress. & 40.3 & \makecell{1.02 \\{\scriptsize (0.31, 0.61, 0.10)}} & 35.2 & 29.1 & \makecell{1.12 \\{\scriptsize (0.06, 0.98, 0.08)}} & 45.9 \\
    Context + Regress. & 39.8 & \makecell{1.81 \\{\scriptsize (0.78, 0.85, 0.18)}} & 40.5 & 31.8 & \makecell{1.18 \\{\scriptsize (0.16, 0.89, 0.13)}} & 46.8 \\
    \hline
  \end{tabular}
  \caption{\label{two_factors}
    Combination of two factors. Roughly sorted by contribution. The parentheses under each LLM API cost report the three LLM API cost components: (input cost of oracle signal contexts, input cost of agent observations, LLM output cost). Different signal injection orders are tested with no significant performance difference (<2\%), so we do not go into detail regarding the differences between the various injection orders.
  }
\end{table*}

\clearpage

\begin{table*}
  \centering
    \begin{tabular}{ccccccc}
    \hline
    & \textbf{Regression} & \textbf{Location} & \textbf{API} & \textbf{Context} & \textbf{Reproduction} \\
    \hline
    \multicolumn{6}{c}{\textit{SWE-bench-Verified: GPT-5-Thinking-Medium + GPT-4o}} \\
    Success gain & +4\% & +12\% & +8\% & +14\% & \textbf{+26\%} \\
    Stage 1 \$ / Step & 0.87\$ / 33.0 & 0.66\$ / 32.2 & 0.89\$ / 32.9 & 0.87\$ / 32.4 & \textbf{0.39\$ / 17.9} \\
    Stage 2 \$ / Step & 1.65\$ / 40.6 & \textbf{0.85\$ / 36.8} & 1.69\$ / 40.9  & 1.51\$ / 41.6 & 1.74\$ / 36.7 \\
    \hline
    \multicolumn{6}{c}{\textit{SWE-bench-Live: Claude-4.6-Sonnet + GPT-5-Thinking-Medium}} \\
    Success gain & +4\% & +8\% & +14\% & +14\% & \textbf{+24\%} \\
    Stage 1 \$/Step & 1.02\$ / 49.2 & \textbf{0.69\$ / 35.5} & 1.09\$ / 53.5 & 0.97\$ / 45.6 & 0.78\$ / 36.4 \\
    Stage 2 \$/Step & 0.68\$/ 32.6 & 0.56\$ / 30.4 & 0.64\$ / 31.5 & 0.72\$ / 33.6 & \textbf{0.47\$ / 24.0} \\
    \hline
    \multicolumn{6}{c}{\textit{SWE-bench-Pro/Python: Claude-4.6-Sonnet + GPT-5-Thinking-Medium}} \\
    Success gain & +2\% & +6\% & +11\% & +9\% & \textbf{+28\%} \\
    Stage 1 \$/Step & 1.15\$ / 55.0 & \textbf{0.82\$ / 35.8} & 1.21\$ / 56.3 & 1.11\$ / 48.1 & 0.94\$ / 39.8 \\
    Stage 2 \$/Step & 0.88\$ / 39.1 & 0.89\$ / 40.1 & 0.89\$ / 39.0 & 0.92\$ / 38.6 & \textbf{0.47\$ / 19.5} \\
    \hline
    \multicolumn{6}{c}{\textit{SWE-bench-Pro/Go: Claude-4.6-Sonnet + GPT-5-Thinking-Medium}} \\
    Success gain & +2\% & +6\% & +12\% & +10\% & \textbf{+19\%} \\
    Stage 1 \$/Step & 1.18\$ / 45.6 & \textbf{0.89\$ / 39.2} & 1.24\$ / 57.6 & 1.21\$ / 49.7 & 1.10\$ / 40.5 \\
    Stage 2 \$/Step & 1.28\$ / 46.6 & 1.26\$ / 56.1 & 1.12\$ / 44.7 & 1.26\$ / 48.8 & \textbf{0.75\$ / 30.4} \\
    \hline
  \end{tabular}
  \caption{\label{validation_cost}
    Details of validation experiment result, including performance gain for each factor compared to the weaker model baseline and average cost of the two stages. Labels are formatted as \textit{Benchmark: (strong model) + (weak model)}.
  }
\end{table*}

\clearpage

\section{Examples of Extracted Factor}
\label{example_signal}

\subsection{Regression Test}

\begin{lstlisting}[language=json]
{
    "instance_id": "verified/django__django-10999",
    "command": "python ./tests/run.py --verbosity 2 --settings=test_pgsql --parallel 1 utils_tests.test_dateparse",
    "PASS_TO_PASS": [
        "test_parse_date (utils_tests.test_dateparse.DateParseTests)",
        "test_parse_datetime (utils_tests.test_dateparse.DateParseTests)",
        "test_parse_time (utils_tests.test_dateparse.DateParseTests)",
        "test_days (utils_tests.test_dateparse.DurationParseTests)", 
        "test_fractions_of_seconds (utils_tests.test_dateparse.DurationParseTests)", 
        "test_hours_minutes_seconds (utils_tests.test_dateparse.DurationParseTests)", 
        "test_iso_8601 (utils_tests.test_dateparse.DurationParseTests)", 
        "test_minutes_seconds (utils_tests.test_dateparse.DurationParseTests)", 
        "test_parse_python_format (utils_tests.test_dateparse.DurationParseTests)", 
        "test_seconds (utils_tests.test_dateparse.DurationParseTests)"
    ]
}
\end{lstlisting}

\subsection{Reproduction Test}

\textbf{Instance\_id:}

verified/django\_\_django-16560.txt

\textbf{Reproduction Test Code:}
\begin{lstlisting}[language=MyPython]
===========================
# constraints.tests.BaseConstraintTests.test_custom_violation_code_message.test_custom_violation_code_message:
1|from unittest import mock
2|
3|from django.core.exceptions import ValidationError
4|from django.db import IntegrityError, connection, models
5|from django.db.models import F
6|from django.db.models.constraints import BaseConstraint, UniqueConstraint
7|from django.db.models.functions import Lower
8|from django.db.transaction import atomic
9|from django.test import SimpleTestCase, TestCase, skipIfDBFeature, skipUnlessDBFeature
10|from django.test.utils import ignore_warnings
11|from django.utils.deprecation import RemovedInDjango60Warning
12|
13|from .models import (
14|    ChildModel,
15|    ChildUniqueConstraintProduct,
16|    Product,
17|    UniqueConstraintConditionProduct,
18|    UniqueConstraintDeferrable,
19|    UniqueConstraintInclude,
20|    UniqueConstraintProduct,
21|)
22|
......
29|class BaseConstraintTests(SimpleTestCase):
......
80|    def test_custom_violation_code_message(self):
81|        c = BaseConstraint(name="base_name", violation_error_code="custom_code")
82|        self.assertEqual(c.violation_error_code, "custom_code")

===========================
# constraints.tests.CheckConstraintTests.test_eq.test_eq:
1|from unittest import mock
2|
3|from django.core.exceptions import ValidationError
4|from django.db import IntegrityError, connection, models
5|from django.db.models import F
6|from django.db.models.constraints import BaseConstraint, UniqueConstraint
7|from django.db.models.functions import Lower
8|from django.db.transaction import atomic
9|from django.test import SimpleTestCase, TestCase, skipIfDBFeature, skipUnlessDBFeature
10|from django.test.utils import ignore_warnings
11|from django.utils.deprecation import RemovedInDjango60Warning
12|
13|from .models import (
14|    ChildModel,
15|    ChildUniqueConstraintProduct,
16|    Product,
17|    UniqueConstraintConditionProduct,
18|    UniqueConstraintDeferrable,
19|    UniqueConstraintInclude,
20|    UniqueConstraintProduct,
21|)
22|
......
120|class CheckConstraintTests(TestCase):
121|    def test_eq(self):
122|        check1 = models.Q(price__gt=models.F("discounted_price"))
123|        check2 = models.Q(price__lt=models.F("discounted_price"))
124|        self.assertEqual(
125|            models.CheckConstraint(check=check1, name="price"),
126|            models.CheckConstraint(check=check1, name="price"),
127|        )
128|        self.assertEqual(models.CheckConstraint(check=check1, name="price"), mock.ANY)
129|        self.assertNotEqual(
130|            models.CheckConstraint(check=check1, name="price"),
131|            models.CheckConstraint(check=check1, name="price2"),
132|        )
133|        self.assertNotEqual(
134|            models.CheckConstraint(check=check1, name="price"),
135|            models.CheckConstraint(check=check2, name="price"),
136|        )
137|        self.assertNotEqual(models.CheckConstraint(check=check1, name="price"), 1)
138|        self.assertNotEqual(
139|            models.CheckConstraint(check=check1, name="price"),
140|            models.CheckConstraint(
141|                check=check1, name="price", violation_error_message="custom error"
142|            ),
143|        )
144|        self.assertNotEqual(
145|            models.CheckConstraint(
146|                check=check1, name="price", violation_error_message="custom error"
147|            ),
148|            models.CheckConstraint(
149|                check=check1, name="price", violation_error_message="other custom error"
150|            ),
151|        )
152|        self.assertEqual(
153|            models.CheckConstraint(
154|                check=check1, name="price", violation_error_message="custom error"
155|            ),
156|            models.CheckConstraint(
157|                check=check1, name="price", violation_error_message="custom error"
158|            ),
159|        )
160|        self.assertNotEqual(
161|            models.CheckConstraint(check=check1, name="price"),
162|            models.CheckConstraint(
163|                check=check1, name="price", violation_error_code="custom_code"
164|            ),
165|        )
166|        self.assertEqual(
167|            models.CheckConstraint(
168|                check=check1, name="price", violation_error_code="custom_code"
169|            ),
170|            models.CheckConstraint(
171|                check=check1, name="price", violation_error_code="custom_code"
172|            ),
173|        )

===========================
'''
<...As the content is too long, parts of this example have been omitted; the actual information provided to the agent is complete...>
'''
===========================
# constraints.tests.UniqueConstraintTests.test_validate_conditon_custom_error.test_validate_conditon_custom_error:
1|from unittest import mock
2|
3|from django.core.exceptions import ValidationError
4|from django.db import IntegrityError, connection, models
5|from django.db.models import F
6|from django.db.models.constraints import BaseConstraint, UniqueConstraint
7|from django.db.models.functions import Lower
8|from django.db.transaction import atomic
9|from django.test import SimpleTestCase, TestCase, skipIfDBFeature, skipUnlessDBFeature
10|from django.test.utils import ignore_warnings
11|from django.utils.deprecation import RemovedInDjango60Warning
12|
13|from .models import (
14|    ChildModel,
15|    ChildUniqueConstraintProduct,
16|    Product,
17|    UniqueConstraintConditionProduct,
18|    UniqueConstraintDeferrable,
19|    UniqueConstraintInclude,
20|    UniqueConstraintProduct,
21|)
22|
......
336|class UniqueConstraintTests(TestCase):
......
811|    @skipUnlessDBFeature("supports_partial_indexes")
812|    def test_validate_conditon_custom_error(self):
813|        p1 = UniqueConstraintConditionProduct.objects.create(name="p1")
814|        constraint = UniqueConstraintConditionProduct._meta.constraints[0]
815|        constraint.violation_error_message = "Custom message"
816|        constraint.violation_error_code = "custom_code"
817|        msg = "Custom message"
818|        with self.assertRaisesMessage(ValidationError, msg) as cm:
819|            constraint.validate(
820|                UniqueConstraintConditionProduct,
821|                UniqueConstraintConditionProduct(name=p1.name, color=None),
822|            )
823|        self.assertEqual(cm.exception.code, "custom_code")

\end{lstlisting}

\textbf{Reproduction Test Command:}
\begin{lstlisting}[language=bash]
./tests/runtests.py \
  --verbosity 2 \
  --settings=test_sqlite \
  --parallel 1 \
  constraints.tests.BaseConstraintTests.test_custom_violation_code_message.test_custom_violation_code_message \
  constraints.tests.BaseConstraintTests.test_deconstruction.test_deconstruction \
  constraints.tests.CheckConstraintTests.test_eq.test_eq \
  constraints.tests.CheckConstraintTests.test_repr_with_violation_error_code.test_repr_with_violation_error_code \
  constraints.tests.CheckConstraintTests.test_validate_custom_error.test_validate_custom_error \
  constraints.tests.UniqueConstraintTests.test_eq.test_eq \
  constraints.tests.UniqueConstraintTests.test_repr_with_violation_error_code.test_repr_with_violation_error_code \
  constraints.tests.UniqueConstraintTests.test_validate_conditon_custom_error.test_validate_conditon_custom_error
\end{lstlisting}

\subsection{Edit Location}

\begin{lstlisting}[language=json]

{
 "instance_id": "pro/golang/future-architect__vuls-54e73c2",
 "detector/detector.go": {
  "line_ranges": [[134, 142], [145, 151], [154, 166]],
  "code_snippets": [
        "......",
        "35|// Detect vulns and fill CVE detailed information",
        "36|func Detect(dbclient DBClient, rs []models.ScanResult, dir string) ([]models.ScanResult, error) {",
        "37|",
        "38|    // Use the same reportedAt for all rs",
        "39|    reportedAt := time.Now()",
        "40|    for i, r := range rs {",
        "41|        if !c.Conf.RefreshCve && !needToRefreshCve(r) {",
        "42|            logging.Log.Info('No need to refresh')",
        "43|            continue",
        "44|        }",
        "45|",
        "46|        if !reuseScannedCves(&r) {",
        "47|            r.ScannedCves = models.VulnInfos{}",
        "48|        }",
        "As the content is too long, parts of this example about the dependencies of the Edit Location in the same file have been omitted; the actual input provided to the agent is complete."
        "114|        rs[i] = r",
        "115|    }",
        "116|",
        "117|    // Overwrite the json file every time to clear the fields specified in config.IgnoredJSONKeys",
        "118|    for _, r := range rs {",
        "119|        if s, ok := c.Conf.Servers[r.ServerName]; ok {",
        "120|            r = r.ClearFields(s.IgnoredJSONKeys)",
        "121|        }",
        "122|        //TODO don't call here",
        "123|        if err := reporter.OverwriteJSONFile(dir, r); err != nil {",
        "124|            return nil, xerrors.Errorf('Failed to write JSON: %w', err)",
        "125|        }",
        "126|    }",
        "127|",
        "128|    if c.Conf.DiffPlus || c.Conf.DiffMinus {",
        "129|        prevs, err := loadPrevious(rs)",
        "130|        if err != nil {",
        "131|            return nil, err",
        "132|        }",
        "133|        rs = diff(rs, prevs, c.Conf.DiffPlus, c.Conf.DiffMinus)",
        "134|    }",
        "135|",
        "136|    for i, r := range rs {",
        "137|        r = r.FilterByCvssOver(c.Conf.CvssScoreOver)",
        "138|        r = r.FilterUnfixed(c.Conf.IgnoreUnfixed)",
        "139|        r = r.FilterInactiveWordPressLibs(c.Conf.WpScan.DetectInactive)",
        "140|",
        "141|        // IgnoreCves",
        "142|        ignoreCves := []string{}",
        "143|        if r.Container.Name == '' {",
        "144|            ignoreCves = c.Conf.Servers[r.ServerName].IgnoreCves",
        "145|        } else if con, ok := c.Conf.Servers[r.ServerName].Containers[r.Container.Name]; ok {",
        "146|            ignoreCves = con.IgnoreCves",
        "147|        }",
        "148|        r = r.FilterIgnoreCves(ignoreCves)",
        "149|",
        "150|        // ignorePkgs",
        "151|        ignorePkgsRegexps := []string{}",
        "152|        if r.Container.Name == '' {",
        "153|            ignorePkgsRegexps = c.Conf.Servers[r.ServerName].IgnorePkgsRegexp",
        "154|        } else if s, ok := c.Conf.Servers[r.ServerName].Containers[r.Container.Name]; ok {",
        "155|            ignorePkgsRegexps = s.IgnorePkgsRegexp",
        "156|        }",
        "157|        r = r.FilterIgnorePkgs(ignorePkgsRegexps)",
        "158|",
        "159|        // IgnoreUnscored",
        "160|        if c.Conf.IgnoreUnscoredCves {",
        "161|            r.ScannedCves = r.ScannedCves.FindScoredVulns()",
        "162|        }",
        "163|",
        "164|        rs[i] = r",
        "165|    }",
        "166|    return rs, nil",
        "167|}",
        "......",
        "169|// DetectPkgCves detects OS pkg cves",
        "170|func DetectPkgCves(dbclient DBClient, r *models.ScanResult) error {",
        "171|    // Pkg Scan",
        "172|    if r.Release != '' {",
        "173|        // OVAL",
        "174|        if err := detectPkgsCvesWithOval(dbclient.OvalDB, r); err != nil {",
        "175|            return xerrors.Errorf('Failed to detect CVE with OVAL: %w', err)",
        "176|        }",
        "177|",
        "178|        // gost",
        "179|        if err := detectPkgsCvesWithGost(dbclient.GostDB, r); err != nil {",
        "180|            return xerrors.Errorf('Failed to detect CVE with gost: %w', err)",
        "181|        }",
        "182|    } else if reuseScannedCves(r) {",
        "183|        logging.Log.Infof('r.Release is empty. Use CVEs as it as.')",
        "184|    } else if r.Family == constant.ServerTypePseudo {",
        "185|        logging.Log.Infof('pseudo type. Skip OVAL and gost detection')",
        "186|    } else {",
        "187|        return xerrors.Errorf('Failed to fill CVEs. r.Release is empty')",
        "188|    }",
        "189|",
        "190|    for i, v := range r.ScannedCves {",
        "191|        for j, p := range v.AffectedPackages {",
        "192|            if p.NotFixedYet && p.FixState == '' {",
        "193|                p.FixState = 'Not fixed yet'",
        "194|                r.ScannedCves[i].AffectedPackages[j] = p",
        "195|            }",
        "196|        }",
        "197|    }",
        "198|",
        "199|    // To keep backward compatibility",
        "200|    // Newer versions use ListenPortStats,",
        "201|    // but older versions of Vuls are set to ListenPorts.",
        "202|    // Set ListenPorts to ListenPortStats to allow newer Vuls to report old results.",
        "203|    for i, pkg := range r.Packages {",
        "204|        for j, proc := range pkg.AffectedProcs {",
        "205|            for _, ipPort := range proc.ListenPorts {",
        "206|                ps, err := models.NewPortStat(ipPort)",
        "207|                if err != nil {",
        "208|                    logging.Log.Warnf('Failed to parse ip:port: %s, err:%+v', ipPort, err)",
        "209|                    continue",
        "210|                }",
        "211|                r.Packages[i].AffectedProcs[j].ListenPortStats = append(",
        "212|                    r.Packages[i].AffectedProcs[j].ListenPortStats, *ps)",
        "213|            }",
        "214|        }",
        "215|    }",
        "216|",
        "217|    return nil",
        "218|}",
        "......",
  ]
 },
 "detector/wordpress.go": {
  "line_ranges": [[110, 116]],
  "code_snippets": [
        "......",
        "51|// DetectWordPressCves access to wpscan and fetch scurity alerts and then set to the given ScanResult.",
        "52|// https://wpscan.com/",
        "53|func detectWordPressCves(r *models.ScanResult, cnf *c.WpScanConf) (int, error) {",
        "54|    if len(r.WordPressPackages) == 0 {",
        "55|            return 0, nil",
        "56|    }",
        "57|    // Core",
        "58|    ver := strings.Replace(r.WordPressPackages.CoreVersion(), '.', '', -1)",
        "59|    if ver == '' {",
        "60|            return 0, errof.New(errof.ErrFailedToAccessWpScan,",
        "61|                    fmt.Sprintf('Failed to get WordPress core version.'))",
        "62|    }",
        "63|    url := fmt.Sprintf('https://wpscan.com/api/v3/wordpresses/%s', ver)",
        "64|    wpVinfos, err := wpscan(url, ver, cnf.Token)",
        "65|    if err != nil {",
        "66|            return 0, err",
        "67|    }",
        "68|",
        "69|    // Themes",
        "70|    themes := r.WordPressPackages.Themes()",
        "71|    if !cnf.DetectInactive {",
        "72|            themes = removeInactives(themes)",
        "73|    }",
        "74|    for _, p := range themes {",
        "75|            url := fmt.Sprintf('https://wpscan.com/api/v3/themes/%s', p.Name)",
        "76|            candidates, err := wpscan(url, p.Name, cnf.Token)",
        "77|            if err != nil {",
        "78|                    return 0, err",
        "79|            }",
        "80|            vulns := detect(p, candidates)",
        "81|            wpVinfos = append(wpVinfos, vulns...)",
        "82|    }",
        "83|",
        "84|    // Plugins",
        "85|    plugins := r.WordPressPackages.Plugins()",
        "86|    if !cnf.DetectInactive {",
        "87|            plugins = removeInactives(plugins)",
        "88|    }",
        "89|    for _, p := range plugins {",
        "90|            url := fmt.Sprintf('https://wpscan.com/api/v3/plugins/%s', p.Name)",
        "91|            candidates, err := wpscan(url, p.Name, cnf.Token)",
        "92|            if err != nil {",
        "93|                    return 0, err",
        "94|            }",
        "95|            vulns := detect(p, candidates)",
        "96|            wpVinfos = append(wpVinfos, vulns...)",
        "97|    }",
        "98|",
        "99|    for _, wpVinfo := range wpVinfos {",
        "100|           if vinfo, ok := r.ScannedCves[wpVinfo.CveID]; ok {",
        "101|                   vinfo.CveContents[models.WpScan] = wpVinfo.CveContents[models.WpScan]",
        "102|                   vinfo.VulnType = wpVinfo.VulnType",
        "103|                   vinfo.Confidences = append(vinfo.Confidences, wpVinfo.Confidences...)",
        "104|                   vinfo.WpPackageFixStats = append(vinfo.WpPackageFixStats, wpVinfo.WpPackageFixStats...)",
        "105|                   r.ScannedCves[wpVinfo.CveID] = vinfo",
        "106|           } else {",
        "107|                   r.ScannedCves[wpVinfo.CveID] = wpVinfo",
        "108|           }",
        "109|   }",
        "110|   return len(wpVinfos), nil",
        "111|}",
        "112|",
        "113|func wpscan(url, name, token string) (vinfos []models.VulnInfo, err error) {",
        "114|   body, err := httpRequest(url, token)",
        "115|   if err != nil {",
        "116|           return nil, err",
        "117|   }",
        "118|   if body == '' {",
        "119|           logging.Log.Debugf('wpscan.com response body is empty. URL: %s', url)",
        "120|   }",
        "121|   return convertToVinfos(name, body)",
        "122|}",
        "......",
        "124|func detect(installed models.WpPackage, candidates []models.VulnInfo) (vulns []models.VulnInfo) {",
        "125|   for _, v := range candidates {",
        "126|           for _, fixstat := range v.WpPackageFixStats {",
        "127|                   ok, err := match(installed.Version, fixstat.FixedIn)",
        "128|                   if err != nil {",
        "129|                           logging.Log.Warnf('Failed to compare versions %s installed: %s, fixedIn: %s, v: %+v',",
        "130|                                   installed.Name, installed.Version, fixstat.FixedIn, v)",
        "131|                           // continue scanning",
        "132|                           continue",
        "133|                   }",
        "134|                   if ok {",
        "135|                           vulns = append(vulns, v)",
        "136|                           logging.Log.Debugf('Affected: %s installed: %s, fixedIn: %s',",
        "137|                                   installed.Name, installed.Version, fixstat.FixedIn)",
        "138|                   } else {",
        "139|                           logging.Log.Debugf('Not affected: %s : %s, fixedIn: %s',",
        "140|                                   installed.Name, installed.Version, fixstat.FixedIn)",
        "141|                   }",
        "142|           }",
        "143|   }",
        "144|   return",
        "145|}",
        "......",
  ]
 }
}

\end{lstlisting}

\subsection{API Usage}

\begin{lstlisting}[language=json]
{
    "instance_id": "sphinx-doc__sphinx-7462",
    "sphinx/domains/python.py": [  
        {
            "file of api definition": "/testbed/sphinx/domains/python.py",
            "lineno of api definition": [92, 111],
            "api function name": "unparse",
            "api function should be called in this way": "unparse(elem)",
            "the approximate lineno where this api appear in the file you need to modify": 111,
            "api function definition": [
                "def _parse_annotation(annotation: str) -> List[Node]:",
                "    '''Parse type annotation.'''",
                "    def make_xref(text: str) -> addnodes.pending_xref:",
                "        if text == 'None':",
                "            reftype = 'obj'",
                "        else:",
                "            reftype = 'class'",
                "",
                "        return pending_xref('', nodes.Text(text),",
                "                            refdomain='py', reftype=reftype, reftarget=text)",
                "",
                "    def unparse(node: ast.AST) -> List[Node]:",
                "        if isinstance(node, ast.Attribute):",
                "            return [nodes.Text('%s.%s' % (unparse(node.value)[0], node.attr))]",
                "        elif isinstance(node, ast.Expr):",
                "            return unparse(node.value)",
                "        elif isinstance(node, ast.Index):",
                "            return unparse(node.value)",
                "        elif isinstance(node, ast.List):",
                "            result = [addnodes.desc_sig_punctuation('', '[')]  # type: List[Node]",
                "            for elem in node.elts:",
                "                result.extend(unparse(elem))",
                "                result.append(addnodes.desc_sig_punctuation('', ', '))",
                "            result.pop()",
                "            result.append(addnodes.desc_sig_punctuation('', ']'))",
                "            return result",
                "        elif isinstance(node, ast.Module):",
                "            return sum((unparse(e) for e in node.body), [])",
                "        elif isinstance(node, ast.Name):",
                "            return [nodes.Text(node.id)]",
                "        elif isinstance(node, ast.Subscript):",
                "            result = unparse(node.value)",
                "            result.append(addnodes.desc_sig_punctuation('', '['))",
                "            result.extend(unparse(node.slice))",
                "            result.append(addnodes.desc_sig_punctuation('', ']'))",
                "            return result",
                "        elif isinstance(node, ast.Tuple):",
                "            if node.elts:",
                "                result = []",
                "                for elem in node.elts:",
                "                    result.extend(unparse(elem))",
                "                    result.append(addnodes.desc_sig_punctuation('', ', '))",
                "                result.pop()",
                "            else:",
                "                result = [addnodes.desc_sig_punctuation('', '('),",
                "                          addnodes.desc_sig_punctuation('', ')')]",
                "",
                "            return result",
                "        else:",
                "            raise SyntaxError  # unsupported syntax",
                "",
                "    try:",
                "        tree = ast_parse(annotation)",
                "        result = unparse(tree)",
                "        for i, node in enumerate(result):",
                "            if isinstance(node, nodes.Text):",
                "                result[i] = make_xref(str(node))",
                "        return result",
                "    except SyntaxError:",
                "        return [make_xref(annotation)]",
            ]
        },
        {
            "file of api definition": "/opt/miniconda3/envs/testbed/lib/python3.9/site-packages/docutils/nodes.py",
            "lineno of api definition": [692, 692],
            "api function name": "append",
            "api function should be called in this way": "result.append(addnodes.desc_sig_punctuation('', ', '))",
            "the approximate lineno where this api appear in the file you need to modify": 112,
            "api function definition": [
                "class Element(Node):",
                "......",
                "    def append(self, item):",
                "        self.setup_child(item)",
                "        self.children.append(item)",
            ]
        }
    ],
    "sphinx/ext/autodoc/__init__.py": [
        {
            "file of api definition": "/testbed/sphinx/util/inspect.py",
            "lineno of api definition": [
                393,
                398
            ],
            "api function name": "isproperty",
            "api function should be called in this way": "inspect.isproperty(member)",
            "the approximate lineno where this api appear in the file you need to modify": 2665,
                "api function definition": [
                "def isproperty(obj: Any) -> bool:",
                "    '''Check if the object is property.'''",
                "    if sys.version_info >= (3, 8):",
                "        from functools import cached_property  # cached_property is available since py3.8",
                "        if isinstance(obj, cached_property):",
                "            return True",
                "",
                "    return isinstance(obj, property)",
            ]
        },
        {
            "file of api definition": "/testbed/sphinx/events.py",
            "lineno of api definition": [
                89,
                89
            ],
            "api function name": "emit",
            "api function should be called in this way": "self.env.app.events.emit(\"source-read\", docname, arg)",
            "the approximate lineno where this api appear in the file you need to modify": 392,
            "api function definition": [
                "class EventManager:",
                "......",
                "    def emit(self, name: str, *args: Any,",
                "             allowed_exceptions: tuple[type[Exception], ...] = ()) -> list:",
                "        '''Emit a Sphinx event.'''",
                "",
                "        # not every object likes to be repr()'d (think",
                "        # random stuff coming via autodoc)",
                "        with contextlib.suppress(Exception):",
                "            logger.debug('[app] emitting event: %r%s', name, repr(args)[:100])",
                "",
                "        results = []",
                "        listeners = sorted(self.listeners[name], key=attrgetter('priority'))",
                "        for listener in listeners:",
                "            try:",
                "                results.append(listener.handler(self.app, *args))",
                "            except allowed_exceptions:",
                "                # pass through the errors specified as *allowed_exceptions*",
                "                raise",
                "            except SphinxError:",
                "                raise",
                "            except Exception as exc:",
                "                if self.app.pdb:",
                "                    # Just pass through the error, so that it can be debugged.",
                "                    raise",
                "                modname = safe_getattr(listener.handler, '__module__', None)",
                "                raise ExtensionError(__('Handler %r for event %r threw an exception') %",
                "                                     (listener.handler, name), exc, modname=modname) from exc",
                "        return results",
            ]
        }
    ]
}

\end{lstlisting}

\subsection{Execution Context}
\textbf{instance\_id: } live/python-attrs\_\_attrs-1319
\\
\textbf{Category: }Native Error Stack Trace with Error Message: SyntaxError("invalid syntax")

\textbf{Stack: }
\begin{lstlisting}[language=json]
[
     [
      "/usr/local/lib/python3.11/site-packages/_pytest/runner.py",
      242,
      "<lambda>",
      []
     ],
     [
      "/usr/local/lib/python3.11/site-packages/_pytest/threadexception.py",
      92,
      "pytest_runtest_call",
      []
     ],
      "/usr/local/lib/python3.11/site-packages/_pytest/python.py",
      159,
      "pytest_pyfunc_call",
      []
     ],
      "/testbed/tests/test_make.py",
      704,
      "test_pre_init_kw_only_work_with_defaults",
      []
     ],
     [
      "/testbed/src/attr/_next_gen.py",
      402,
      "define",
      [
        "387:                break",
        "388:",
        "389:        if auto_attribs is not None:",
        "390:            return do_it(cls, auto_attribs)",
        "391:",
        "392:        try:",
        "393:            return do_it(cls, True)",
        "394:        except UnannotatedAttributeError:",
        "395:            return do_it(cls, False)",
        "396:",
        "397:    # maybe_cls's type depends on the usage of the decorator.  It's a class",
        "398:    # if it's used as `@attrs` but `None` if used as `@attrs()`.",
        "399:    if maybe_cls is None:",
        "400:        return wrap",
        "401:",
        "402:    return wrap(maybe_cls)",
        "403:",
        "404:",
        "405:mutable = define",
        "406:frozen = partial(define, frozen=True, on_setattr=None)",
        "407:",
        "408:",
        "409:def field(",
        "410:    *,",
        "411:    default=NOTHING,",
        "412:    validator=None,",
        "413:    repr=True,",
        "414:    hash=None,",
        "415:    init=True,",
        "416:    metadata=None,",
        "417:    type=None,",
      ]
     ],
     [
      "/testbed/src/attr/_next_gen.py",
      393,
      "wrap",
      [
        "378:        # However, if we subclass a frozen class, we inherit the immutability",
        "379:        # and disable on_setattr.",
        "380:        for base_cls in cls.__bases__:",
        "381:            if base_cls.__setattr__ is _frozen_setattrs:",
        "382:                if had_on_setattr:",
        "383:                    msg = 'Frozen classes can't use on_setattr (frozen-ness was inherited).'",
        "384:                    raise ValueError(msg)",
        "385:",
        "386:                on_setattr = setters.NO_OP",
        "387:                break",
        "388:",
        "389:        if auto_attribs is not None:",
        "390:            return do_it(cls, auto_attribs)",
        "391:",
        "392:        try:",
        "393:            return do_it(cls, True)",
        "394:        except UnannotatedAttributeError:",
        "395:            return do_it(cls, False)",
        "396:",
        "397:    # maybe_cls's type depends on the usage of the decorator.  It's a class",
        "398:    # if it's used as `@attrs` but `None` if used as `@attrs()`.",
        "399:    if maybe_cls is None:",
        "400:        return wrap",
        "401:",
        "402:    return wrap(maybe_cls)",
        "403:",
        "404:",
        "405:mutable = define",
        "406:frozen = partial(define, frozen=True, on_setattr=None)",
      ]
     ],
     [
      "/testbed/src/attr/_next_gen.py",
      339,
      "do_it",
      [
        "324:        - *slots=True*",
        "325:",
        "326:          Usually, this has only upsides and few visible effects in everyday",
        "327:          programming. But it *can* lead to some surprising behaviors, so",
        "328:          please make sure to read :term:`slotted classes`.",
        "329:",
        "330:        - *auto_exc=True*",
        "331:        - *auto_detect=True*",
        "332:        - *order=False*",
        "333:        - Some options that were only relevant on Python 2 or were kept around",
        "334:          for backwards-compatibility have been removed.",
        "335:",
        "336:    '''",
        "337:",
        "338:    def do_it(cls, auto_attribs):",
        "339:        return attrs(",
        "340:            maybe_cls=cls,",
        "341:            these=these,",
        "342:            repr=repr,",
        "343:            hash=hash,",
        "344:            unsafe_hash=unsafe_hash,",
        "345:            init=init,",
        "346:            slots=slots,",
        "347:            frozen=frozen,",
        "348:            weakref_slot=weakref_slot,",
        "349:            str=str,",
        "350:            auto_attribs=auto_attribs,",
        "351:            kw_only=kw_only,",
        "352:            cache_hash=cache_hash,",
        "353:            auto_exc=auto_exc,",
        "354:            eq=eq,",
      ]
     ],
     [
      "/testbed/src/attr/_make.py",
      1403,
      "attrs",
      [
        "1388:",
        "1389:        if (",
        "1390:            PY_3_10_PLUS",
        "1391:            and match_args",
        "1392:            and not _has_own_attribute(cls, '__match_args__')",
        "1393:        ):",
        "1394:            builder.add_match_args()",
        "1395:",
        "1396:        return builder.build_class()",
        "1397:",
        "1398:    # maybe_cls's type depends on the usage of the decorator.  It's a class",
        "1399:    # if it's used as `@attrs` but `None` if used as `@attrs()`.",
        "1400:    if maybe_cls is None:",
        "1401:        return wrap",
        "1402:",
        "1403:    return wrap(maybe_cls)",
        "1404:",
        "1405:",
        "1406:_attrs = attrs",
        "1407:'''",
        "1408:Internal alias so we can use it in functions that take an argument called",
        "1409:*attrs*.",
        "1410:'''",
        "1411:",
        "1412:",
        "1413:def _has_frozen_base_class(cls):",
        "1414:    '''",
        "1415:    Check whether *cls* has a frozen ancestor by looking at its",
        "1416:    __setattr__.",
        "1417:    '''",
        "1418:    return cls.__setattr__ is _frozen_setattrs",
      ]
     ],
     [
      "/testbed/src/attr/_make.py",
      1382,
      "wrap",
      ["<...As the content is too long, parts of this example have been omitted; the actual information provided to the agent is complete...>"]
     ],
     [
      "/testbed/src/attr/_make.py",
      972,
      "add_init",
      ["<...As the content is too long, parts of this example have been omitted; the actual information provided to the agent is complete...>"]
     ],
     [
      "/testbed/src/attr/_make.py",
      1875,
      "_make_init",
      ["<...As the content is too long, parts of this example have been omitted; the actual information provided to the agent is complete...>"]
     ],
     [
      "/testbed/src/attr/_make.py",
      239,
      "_make_method",
      [
        "224:    base_filename = filename",
        "225:    while True:",
        "226:        linecache_tuple = (",
        "227:            len(script),",
        "228:            None,",
        "229:            script.splitlines(True),",
        "230:            filename,",
        "231:        )",
        "232:        old_val = linecache.cache.setdefault(filename, linecache_tuple)",
        "233:        if old_val == linecache_tuple:",
        "234:            break",
        "235:",
        "236:        filename = f'{base_filename[:-1]}-{count}>'",
        "237:        count += 1",
        "238:",
        "239:    _compile_and_eval(script, globs, locs, filename)",
        "240:",
        "241:    return locs[name]",
        "242:",
        "243:",
        "244:def _make_attr_tuple_class(cls_name, attr_names):",
        "245:    '''",
        "246:    Create a tuple subclass to hold `Attribute`s for an `attrs` class.",
        "247:",
        "248:    The subclass is a bare tuple with properties for names.",
        "249:",
        "250:    class MyClassAttributes(tuple):",
        "251:        __slots__ = ()",
        "252:        x = property(itemgetter(0))",
        "253:    '''",
        "254:    attr_class_name = f'{cls_name}Attributes'",
      ]
     ],
     [
      "/testbed/src/attr/_make.py",
      211,
      "_compile_and_eval",
      [
        "196:        kw_only=kw_only,",
        "197:        eq=eq,",
        "198:        eq_key=eq_key,",
        "199:        order=order,",
        "200:        order_key=order_key,",
        "201:        on_setattr=on_setattr,",
        "202:        alias=alias,",
        "203:    )",
        "204:",
        "205:",
        "206:def _compile_and_eval(script, globs, locs=None, filename=''):",
        "207:    '''",
        "208:    Evaluate the script with the given global (globs) and local (locs)",
        "209:    variables.",
        "210:    '''",
        "211:    bytecode = compile(script, filename, 'exec')",
        "212:    eval(bytecode, globs, locs)",
        "213:",
        "214:",
        "215:def _make_method(name, script, filename, globs, locals=None):",
        "216:    '''",
        "217:    Create the method with the script given and return the method object.",
        "218:    '''",
        "219:    locs = {} if locals is None else locals",
        "220:",
        "221:    # In order of debuggers like PDB being able to step through the code,",
        "222:    # we add a fake linecache entry.",
        "223:    count = 1",
        "224:    base_filename = filename",
        "225:    while True:",
        "226:        linecache_tuple = (",
      ]
    ]
]
\end{lstlisting}

\section{Prompts for SWE-agent to Inject Oracle Information}
\label{sec:appendixB}

This section provides how we inject oracle information of five factors into user prompt to run SWE-agent.

\begin{lstlisting}[language=MyPython]
# gt is the dict to store ground truth of the five factors.
USER_PROMPT = ""

if edit_location:
    USER_PROMPT += "Your colleague has found all the locations that should be edited to resolve the issue:\n"
    for location, code in gt["location_content"].items():
        USER_PROMPT += f"#{location}: \n{code}\n\n"
    USER_PROMPT += f"""
Your colleague has determined that only these locations should be edited.
You must edit all of these locations one by one.
You must not edit any locations not listed here.
As your colleagues have done the localization step for you, you don't need to do this step again. You can of course view other contexts relevant to these buggy locations to understand the source codes better. But locations not listed here should not be edited!
"""
    
if reproduction:
    self.sandbox_environment.communicate(f"""git apply - <<'NEW_PATCH'\n{gt["test_patch"]}\nNEW_PATCH""")
    USER_PROMPT = "Your colleague has written the correct reproduction tests in the test folder and has figured out how to run regression tests in the repository.\n"
    USER_PROMPT += "The testcases to reproduce the bugs are:\n"
    for test_name, code in gt["F2P_content"].items():
        USER_PROMPT += f"# {test_name}:\n{code}\n"
    USER_PROMPT += f"The command to run these reproduction testcases is: \n{gt["f2p_cmd"]}\n\n"
    USER_PROMPT += f"The list of reproduction testcases that you must pass: \n{gt["FAIL_TO_PASS"]}\n\n"
    USER_PROMPT += "As your colleague has written the reproduction tests for you, you do not need to do this again. After you modified the source codes, please run reproduction tests to validate your fix. You must pass all the reproduction tests before you submit.\n"

if regression:
    USER_PROMPT += f"The command to run regression tests is: \n{test_cmd} {write_to_file}\n\n"
    USER_PROMPT += f"The list of regression testcases that you must pass: \n{condense_too_long_P2P(gt["PASS_TO_PASS"])}\n\n"
    USER_PROMPT += "As your colleague has found the regression tests for you, you do not need to do this again. After you modified the source codes, please run regression tests to validate your fix. You must pass all the regression tests specified above before you submit.\n"


if context:
    USER_PROMPT = f"""
Hints: Your colleague has provided the error stack traces from running reproduction testcaces related to this problem.
Each layer of trace is in the format of (file path, line number, function name, source code)
{gt["error_stack"]}
The error stack traces can be some hints to help you locate the locations to be edited, and understand the execution contexts which affect the locations to be edited or your edits would affect.
"""

if api:
    USER_PROMPT = f"""
Hints: Your colleague has listed the functions you need to use to edit the source code.
The function information format is a mapping: Dict [ "the path of the file you need to edit" : List [ "the information of each function you need to use to edit this file" ] ]
{gt["api"]}
You need to use ALL of the function calls listed above. You need to use the exactly same arguments in the "api function should be called in this way" field in each piece of function information to use each function.
Some of the function information has best-guess function definition from our development tools, but may not be accurate, so you need to judge and find the correct function definition if necessary.
"""

if reproduction and regression and location:
    user_requirement = ("Following your colleagues' contributions, your next steps should be:\n"
               "1) Explore the source codes if you want to understand the contexts of the buggy locations or testcases better. But remember that all the locations that need to edited have been listed above.\n"
               "2) Edit all the locations specified by your colleagues to resolve the issue.\n"
               "3) Run reproduction and regression tests to validate your fix. If any of the testcases required above fails, go back to explore and edit again.\n"
               "4) If you pass both reproduction and regression tests, submit.\n"
            )
elif reproduction and regression:
    USER_PROMPT = """
Following your colleagues' contributions, your next steps should be:
1) As a first step, it might be a good idea to find and read code relevant to the <pr_description>. 
2) Edit the codes to resolve the issue.
3) Run reproduction and regression tests to validate your fix. If any of the testcases required above fails, go back to explore and edit again.
4) If you pass both reproduction and regression tests, submit.
"""
            )
elif location:
    USER_PROMPT = """
Following your colleagues' contributions, your next steps should be:
1) Explore the source codes if you want to understand the contexts of the buggy locations or testcases better. But remember that all the locations that need to edited have been listed above.
2) Create a script to reproduce the problem and execute it using the bash tool to confirm the problem.
3) Edit all the locations specified by your colleagues to resolve the issue.\n"
4) Run your reproduction script to validate your fix. Due to time limit, do not run the regression tests in the repository. If the reproduction script still fails, go back to explore and edit again.
5) If you pass your reproduction script, submit.
"""

elif reproduction:
    USER_PROMPT = """
Following your colleagues' contributions, your next steps should be:\n"
1) As a first step, it might be a good idea to find and read code relevant to the <pr_description>.
2) Edit the codes to resolve the issue.
3) Run reproduction tests to validate your fix. Due to time limit, do not run the regression tests in the repository. If any of the reproduction testcases listed above fails, go back to explore and edit again.
4) If you pass ALL the reproduction tests, submit.
"""
            )
elif regression:
    USER_PROMPT = """
Follow these steps to resolve the issue:
1. As a first step, it might be a good idea to find and read code relevant to the <pr_description>.
2. Create a script to reproduce the problem and execute it using the bash tool to confirm the problem.
3. Edit the sourcecode of the repo to resolve the issue.
4. Rerun your reproduce script to confirm that the error is fixed. Run the regression tests found by your colleague to check that your edits have not broken anything else. If your reproduce script or any of the regression testcases required above fail, go back to explore and edit again.
5. If you pass both your reproduce script and all the regression tests found by your colleague, submit your answer.
"""
else:
    USER_PROMPT = """
Follow these steps to resolve the issue:
1. As a first step, it might be a good idea to find and read code relevant to the <pr_description>
2. Create a script to reproduce the error and execute it using the bash tool to confirm the error
3. Edit the sourcecode of the repo to resolve the issue
4. Rerun your reproduce script and confirm that the error is fixed!
5. Think about edgecases and make sure your fix handles them as well
Your thinking should be thorough and so it's fine if it's very long.
"""

\end{lstlisting}

\section{Agent Prompt of Validation Experiment Stage 1 (Signal Extraction)}
\label{val_prompt}

\begin{lstlisting}[language=MyPython]
if reproduction:
    if lang == "python":
        target_file = "/testbed/reproduction.py"
    elif lang == "go":
        target_file = "/testbed/reproduction_test.go"
else:
    target_file = "/testbed/answer.json"

if location:
    message = f"""
You need to find the locations that need to be edited to resolve the issue.
You don't need to edit the code.
You only need to find the locations that need to be edited.
You answer should be written into /testbed/answer.json
We will check the correctness of your answer by reading /testbed/answer.json
In /testbed/answer.json, your answer format should be like:

<Edit Location example in Appendix `Examples of Extracted Factor`>

Once you have completed /testbed/answer.json, submit.
"""

if regression:
    message = f"""
You need to find the correct command to run original regression tests in the repository, and parse the test output to record passed testcases and failed testcases.
For many repos the regression command is not trivial -- "pytest . / gotest . " does not resolve everything.
You need to read the repo documents to figure out the correct way to run regression tests -- the test command, the evironment variables to set...
Usually you don't need to run all tests in the repository -- it's too time-consuming. It is recommended to only test codes related to the issue description in your test command.
After regression test, direct test output to log file and parse the test output to record passed and failed testcases. If the test framework allows JSON / XML structured output, use it to help log parsing.
You answer should be written into /testbed/answer.json
We will check the correctness of your answer by reading /testbed/answer.json
In /testbed/answer.json, your answer format should be like:

<Regression Test example in Appendix `Examples of Extracted Factor`>

You don't need to edit the source codes. 
Once you have completed /testbed/answer.json, submit.
"""

if reproduction:
    if lang == "python":
        target_func = ""
        cmd = "cd /testbed/ && python reproduction.py"
    elif lang == "go":
        target_func = "The entry of your test file should be func TestReproduce(t *testing.T) "
        cmd = "cd /testbed/ && go test -run TestReproduce -v"
    message = f"""
You need to write a reproduction test file to help your colleague reproduce the issue.
Your reproduction file, when executed, should exactly reveal the issue in the issue description.
Your reproduction file should fail before the issue is resolved and should pass after the issue is correctly fixed.
Your should write multiple testcases in the reproduction file to cover all corner cases to ensure all cases are resolved correclty.
Your reproduction file should be written in {target_file} using `str_replace_editor create`
{target_func}
We will only extract contents in {target_file} to check the correctness of your reproduction. Don't write anything elsewhere.
Your reproduction test file should be run with `{cmd}`
After you write the reproduction test file, run it with `{cmd}` to confirm it can reproduce the issue in the description.
You don't need to edit the source codes. 
Once you have run the {target_file} to confirm it can reproduce the issue, submit.
    """
    
if api:
    message = f"""
You need to find:
    1) all the API functions that need to be used to edit the code to resolve the issue;
    2) the important API functions that are used around the locations to be edited which can help understand how to edit the code.
These API functions are crucial for understanding how to edit the code and avoiding implementing existing utilities again.
You need to determine where the API functions are used / will be used for editing the source code, and find the definition of these API functions.
You may need to locate the suspicous locations that might need to be edited to resolve the issue first, before you can discover the necessary API functions to help understand / edit these locations.
You answer should be written into /testbed/answer.json
We will check the correctness of your answer by reading /testbed/answer.json
In /testbed/answer.json, your answer format should be like:

<api example in Appendix `Examples of Extracted Factor`>

You don't need to edit the source codes. 
Once you have completed /testbed/answer.json, submit.
"""

if context:
    message = f"""
You need to find the locations relevant to the issue description, not limited to the locations to be edited, for your colleague.
Organize the relations of potentially relevant locations you find into function call stacks.
You answer should be written into /testbed/answer.json
We will check the correctness of your answer by reading /testbed/answer.json
In /testbed/answer.json, your answer format should be like:
[
    [
        ["file path 1", line number, "function name a", "code snippet"],
        ["file path 2", line number, "function name b", "code snippet"],
        ["file path 3", line number, "function name c", "code snippet"],
    ],
    [ stack 2 ... ],
    ...
]
where a calls b, and b calls c.

For example,

<context example in Appendix `Examples of Extracted Factor`>

Requirements:
    (1) Include both the suspicous locations that may need to be edited, and the code contexts related to these locations to be edited, which may affect the locations to be edited or the edited locations may affect, in your final answer.
    (2) Some relevant locations could be found by error stack trace of running issue reproduction test code -- this is the most optimistic cases. You can try to use this way first to get the answer stacks. 
    (3) For problems that cannot be located with error stack of running issue reproduction test code, find a) the locations to be edited first and then find b) the functions which call these edit locations and these edit locations call. Organize the function call relations of BOTH a AND b into stacks.
    (4) Organize the relations of potentially relevant locations you find into list of function call stacks.

You don't need to edit the source codes. 
Once you have completed /testbed/answer.json, submit.
"""

\end{lstlisting}
\end{document}